\documentclass[twocolumn]{aastex63}
\usepackage{graphicx}

\submitjournal{PASP}
\shorttitle{Star-Image Centering with Deep Learning}
\shortauthors{Casetti-Dinescu et al.}

\begin{document}


\title{Star-Image Centering with Deep Learning: HST/WFPC2 Images}

\correspondingauthor{Dana I. Casetti-Dinescu}
\email{casettid1@southernct.edu,dana.casetti@gmail.com}

\author[0000-0001-9737-4954]{Dana I. Casetti-Dinescu}
\affiliation{Department of Physics, Southern Connecticut State University, 501 Crescent Street, 
New Haven, CT 06515}
\affiliation{Astronomical Institute of the Romanian Academy, Cutitul de Argint 5, Sector 4, 
Bucharest, Romania}
\author{Terrence M. Girard}
\affiliation{Department of Physics, Southern Connecticut State University, 501 Crescent Street, 
New Haven, CT 06515}
\author[0000-0001-5214-7408]{Roberto Baena-Gall\'{e}}
\affiliation{Universidad Internacional de la Rioja, Avenida de la Paz, 137,26006, Logro\~{n}o, La Rioja, Spain}
\author{Max Martone}
\affiliation{Department of Physics, Southern Connecticut State University, 501 Crescent Street, 
New Haven, CT 06515}
\author{Kate Schwendemann}
\affiliation{Avon High School, Avon, CT 06001}

\begin{abstract}
  A Deep Learning (DL) algorithm is built and tested for
its ability to determine centers of 
star images on HST/WFPC2 exposures, in filters F555W and F814W. 
These archival observations hold great potential for proper-motion studies, 
but the undersampling in the camera's detectors presents challenges for 
conventional centering algorithms. 
Two exquisite data sets of over 600 exposures of the cluster NGC 104 
in these filters are used as a testbed for training and evaluation of the
DL code.
Results indicate a single-measurement standard error of from 8.5 to 11 mpix,
depending on detector and filter.
This compares favorably to the $\sim20$ mpix achieved with the customary
``effective PSF'' centering procedure for WFPC2 images.
Importantly, pixel-phase error is largely eliminated when using the DL method.
The current tests are limited to the central portion of each detector;
in future studies the DL code will be modified to allow for the known
variation of the PSF across the detectors.

\end{abstract}


\section{Introduction \label{sec:intro}}

Deep learning is a class of Machine Learning inspired
by the human brain structure and operation. This approach is based
on layers of representation specialized in detecting features from
the data to be analyzed. Different models are designed depending
on the signal (voice, image, etc.) and the particular
application (segmentation, denoising, etc.).
Specifically, Convolutional Neural Networks (CNNs) are common in visual
imagery and make use of convolutional kernels to create
feature maps at each layer of representation.
Deep learning \citep{lecun2015} has taken over fields such as
natural language processing, medical image analysis, image
restoration and reconstruction, recommendation systems and many others. 

In astronomy this technique was pioneered in problems such
as galaxy classifications and 
galaxy/star discrimination  \citep[e.g.,][]{kim2017}
and is used now in numerous applications 
such as stellar variability and pulsations \citep{Martinez2022},
star-cluster classification \citep{Castro2022},
galactic structure \citep{Dropulic2021},
photometric redshift \citep{Henghes2022}, etc. 
Given the unprecedented amount of data available
for astronomers, this technique is here to stay, 
and will probably be the norm for the foreseeable future.

In astrometry, only timid attempts to utilize machine learning
have been made thus far, with a focus on determining a
PSF model for ground-based, wide-field surveys \citep{herbel2018}. 
In general, these efforts for obtaining an accurate PSF have as objective the measurement of 
the volume under the PSF for photometry, its shape for weak lensing, or to discriminate between 
point sources and fuzzy, extended objects. 
Here, we begin to explore the use of deep learning methods with the specific purpose of measuring 
more accurate centers of point sources, and consequently improve the centering precision
and overall astrometry that can be achieved with CCD detectors.
We will focus on undersampled images such as those taken with the
wide field planetary camera 2 (WFPC2) on the {\it Hubble Space Telescope} (HST)
as a testbed for these techniques.

WFPC2 exposures were taken between 1993 and 2009, the period the camera was active on HST.
These archived images offer a unique time baseline of between $\sim20$ and 30 years
for high-precision, proper-motion studies
that would incorporate either ACS/WFC HST exposures or JWST exposures as second epoch.
Recent astrometric calibration work
for the WFPC2 provided an improved distortion solution
\citep{casettiwfpc2}, and found no other fixed-pattern systematic effects
on length scales smaller than 34 pixels, besides the
well-documented 34th-row problem \citep{anderson1999wfpc2}.
However, on the scale of a single pixel one must deal with an error
known as the pixel-phase error, which is particularly prevalent in
undersampled images. This error is caused by a 
mismatch between the PSF used in the centering algorithm
and the true PSF of a given image, leading to a bias in the derived fractional-pixel
position of the fitted PSF. \citet{and2000}
addressed this
issue by building an {\it effective} PSF (ePSF)
empirically from
observations. However, the ePSF library
developed for the WFPC2
is not sufficient to remove this pixel-phase
bias, particularly when the target
images are taken at different epochs from the
images used to build the ePSF library.
To this end, we explore an entirely
new approach to centering these
undersampled images. We use a deep-learning
technique together
with various training samples of simulated and real data to
address the pixel-phase bias problem.

\section{Formulating the Problem \label{sec:problem}}

The undersampled nature of star images in WFPC2 exposures
poses special problems for astrometric centering.
Traditional techniques, such as fitting an analytic function
to each stellar
profile, deliver centers 
that are biased by as much as 0.05 pixels
(depending on the function used), with centers 
preferentially located to either the centers or the edges of pixels. 
A histogram of the fractional pixel of such centers clearly shows
a non-uniform distribution;
\citet{and2000} called this effect "pixel-phase error". 
To address this problem, and simultaneously improve the derived photometry, 
\citet{and2000} developed the effective PSF (ePSF) method of image fitting.
The ePSF differs from the actual PSF in that it effectively (and implicitly)
contains an integration across the area of each pixel, times the pixel response function.
Thus, in practice,
when determining the amplitude of the ePSF at the location of a specific pixel --
for the purpose of star-profile fitting -- one need simply evaluate the ePSF as opposed to
computing a 2-d numerical integration across the pixel.
The ePSF is constructed empirically from numerous star images extracted from 
a suitable set of repeated, offset exposures.
Typically the ePSF is represented on a super-resolution grid, thus alleviating the effects
of pixel-phase error.
Employing the ePSF technique, \citet{and2000} demonstrated they could
largely eliminate the pixel-phase bias and obtain uncertainty estimates for 
WFPC2 stellar positions 
of the order of 20 mpix, with the residual errors being random in nature. 
However, the existing ePSF library constructed by \citet{and2000}, upon which a practical
application of the technique relies,
is based on a limited set of observations. Furthermore, the
quoted errors of 20 mpix
were determined from the same set of observations used
to build the ePSF library \citep{and2000}. 
Ultimately, the precision of this technique will depend on the extent to which 
it can be applied to exposures having the same (or very similar) 
PSFs to those used to construct the library ePSFs. 

To better illustrate this pixel-phase bias we make use of a unique data
set in the WFPC2 archive.
The set --- PID 8267 Gilliland --- was taken at the core of globular
cluster 47 Tuc (NGC 104) in July 1999.
All exposures are 160 sec, and there are 636 in filter
F555W and 654 in F814W. The offsets
between exposures are small (up to 2 PC pixels, or 1 WF pixel ), and,
while not spatially random,
they do not have a quantized, fixed pattern either.
The small, fractional-pixel offsets
are very helpful in characterizing the pixel-phase bias,
since on the scale of these offsets effects due to differential distortion and the 34th-row
error can effectively be ignored.

The ePSF technique is implemented in the
hst1pass\footnote{While using a 2019 version of the code,
we have checked that the most recent 2022 version
\citep{anderson2022} gives the same results;
the ePSF library is identical in the two versions of the code.} code
\citep{and2000,anderson2022}.
We use this code
to obtain detections, object centers, instrumental magnitudes and
a quality-of-fit parameter (q). We use only objects with q = 0.0001 to 1.0.
At the lower limit (q = 0.0001), the restriction is intended to eliminate
bright stars which are not necessarily saturated, but are deemed not
well centered by the hst1pass algorithm, which fits only the inner
$5 \times 5$ pixels of an object.

\begin{figure} 
\includegraphics[scale=0.45,angle=-90]{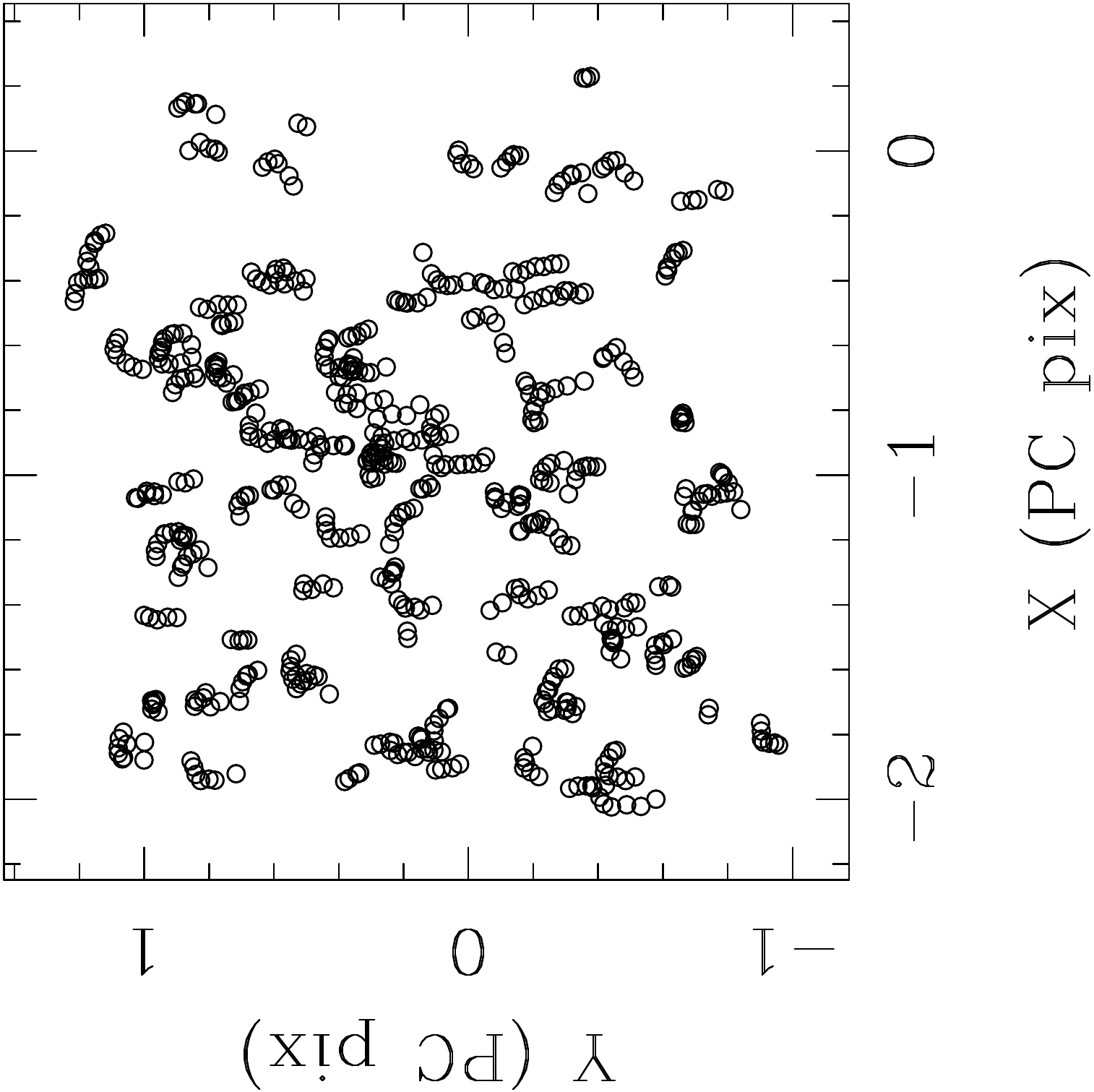}
\caption{Distribution of offsets with respect to the reference
image for the F555W data set.
There are 635 offsets shown; the reference image is at
$(X,Y) = (0., 0.)$}
\label{fig:offsets}
\end{figure}

Precise offsets between the 636 exposures were determined
from the transformation of each exposure into the
first one of the set, which is taken as a reference exposure.
We use a general cubic polynomial transformation, in each coordinate, between the
reference image positions and the 
target image positions. The coefficients (including the offsets)
are determined by an iterative least-squares procedure
with outlier culling. The residuals of this transformation are
the differences between the positions of the
stars in the reference image and the transformed positions of
same stars in the target image. The standard error of the
transformation will be used as a diagnostic tool in what follows.
In Figure \ref{fig:offsets} we show
the distribution of derived offsets in PC-pixels (0.046 ``/pixel) for
filter F555W; filter F814W has a nearly-identical distribution.

\begin{figure} 
\includegraphics[scale=0.45,angle=0]{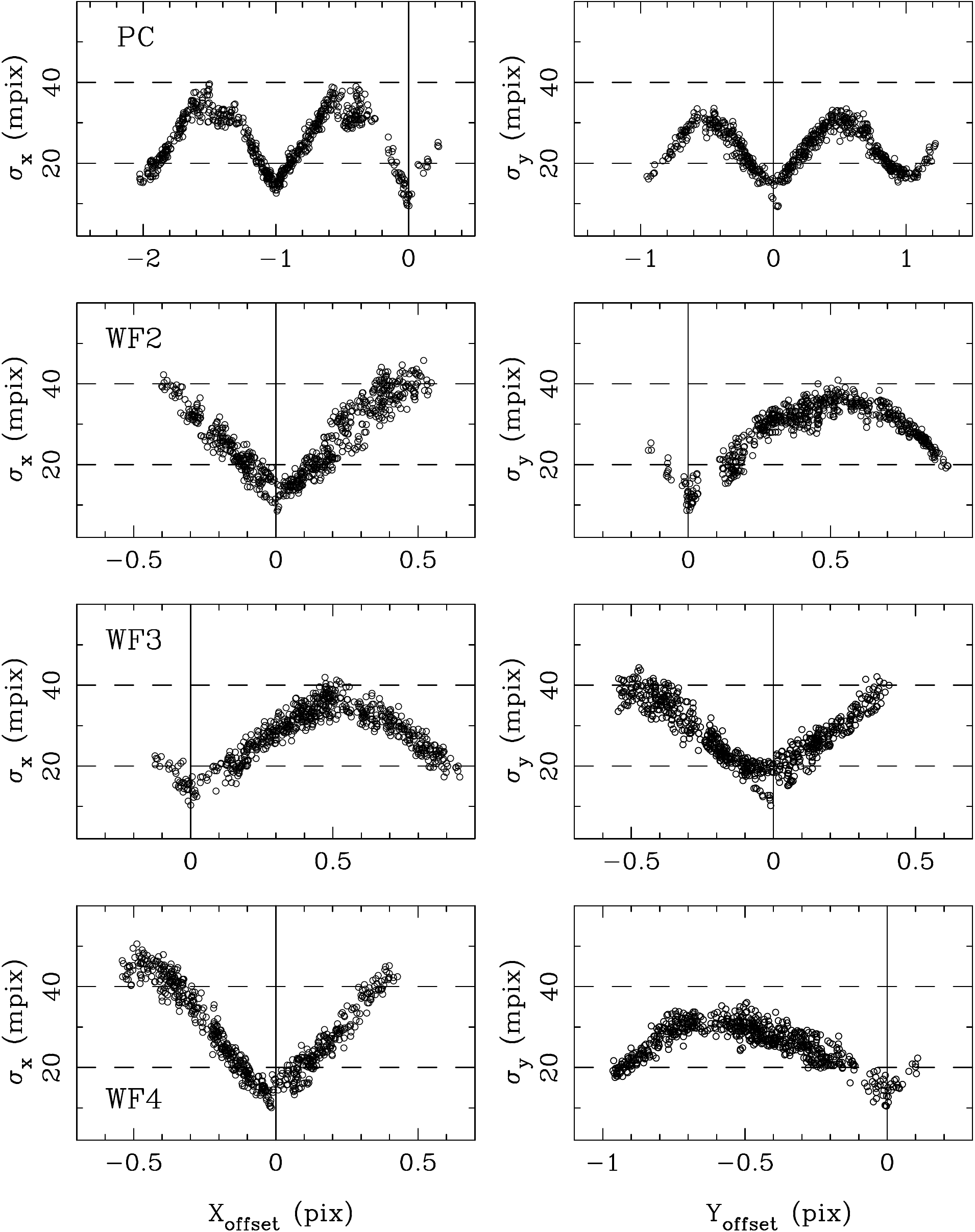}
\caption{Standard error of the transformation of
each exposure's star positions into those of the
reference exposure, as a function of the offset
from the reference exposure.
Star centers are obtained with hst1pass
\citep{and2000,anderson2022}.
We restrict this test to the central area of each chip.
Dashed lines mark the 20- and 40-mpix level.
The vertical line marks the 0-pixel offset. 
The variation of the standard error with 
offset is striking
and demonstrates the presence of the
pixel-phase bias.}
\label{fig:stderror-hst1pass}
\end{figure}

In Figure \ref{fig:stderror-hst1pass} we plot the standard error
of the transformations as a function of offset for both
x and y coordinates, for all exposures. 
Each WFPC2 chip is treated separately.
In the transformations we use the raw positions from hst1pass, uncorrected for
distortion and the 34th-row bias, and restrict ourselves
to the central third of the chip in each coordinate,
i.e., stars in the central 1/9th area of each chip.
The hst1pass code employs a library of ePSFs specifically
constructed for the WFPC2. For each chip a
$3 \times 3$ x-y grid of
ePSFs is provided to account for the variability of the
ePSF across each chip. Since in the current work we are solely
concerned with better modeling the undersampling effect
on stars' centers, we will focus only
on the central 1/9 part of the chip,
which is assumed to be
characterized by a constant ePSF. For this reason,
throughout the paper, when considering real data,
we will restrict the
analysis to the central part of each chip.

\begin{figure} 
\includegraphics[scale=0.40,angle=0]{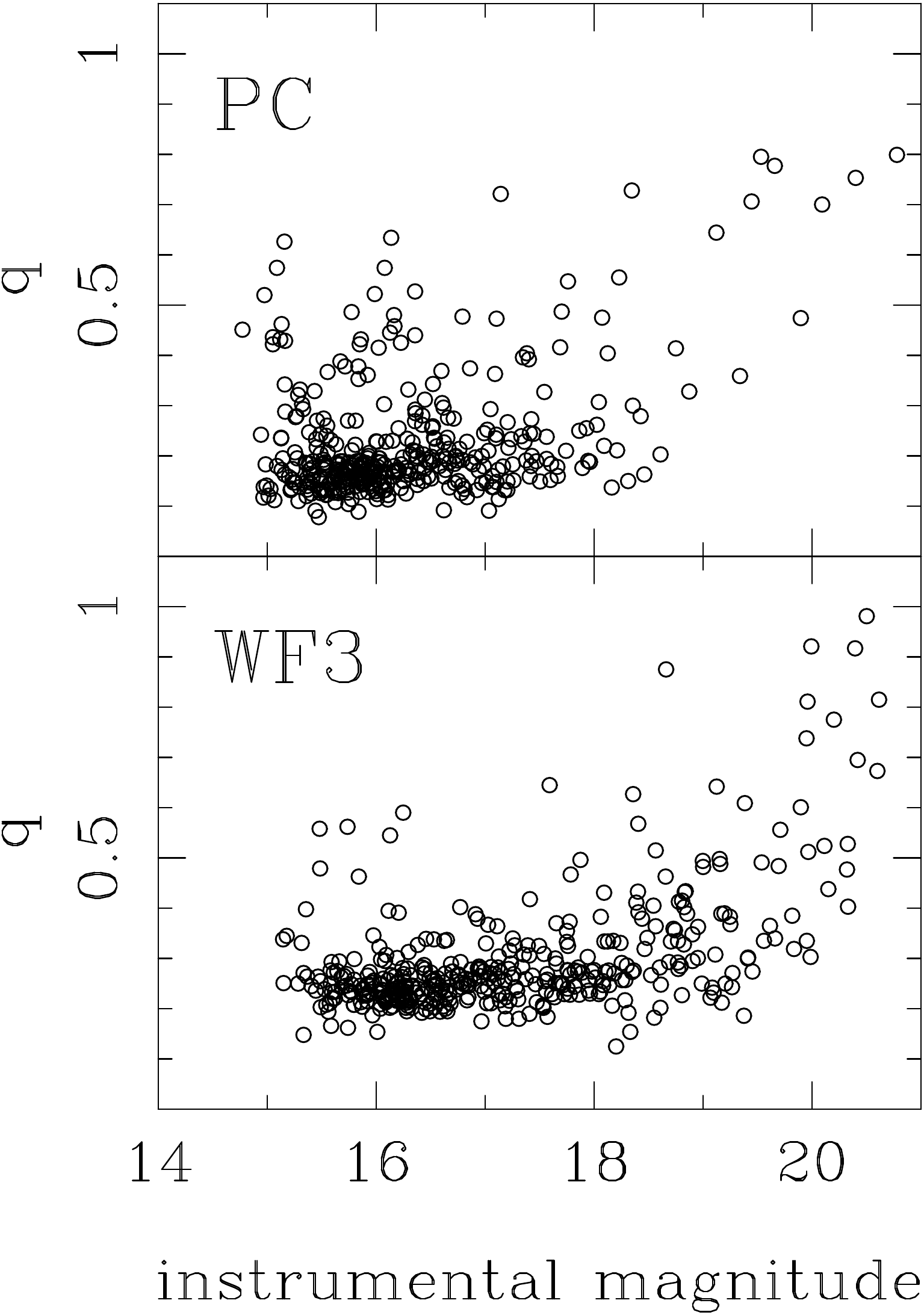}
\caption{Quality-of-fit, q, as a function of magnitude for one representative exposure
  of the NGC 104 data set. Only 
objects within the central $11\%$ area of the chip are considered.}
\label{fig:q-vs-mag}
\end{figure}

Figure \ref{fig:stderror-hst1pass} shows a striking
dependence of the standard error of the transformation
with image offset from the reference image. When the offset
is an integer pixel, the standard error is small, as all stars
fall on the same fractional part of a pixel as in the reference exposure.
As the offset departs from an integer pixel, the standard
error increases to a maximum at (modulo) half-pixel offsets.
Such a dependence is explained by the presence
of pixel-phase bias. That is, the fractional part of the derived star centers
is not randomly distributed; rather, it prefers the
center or the edge of a pixel. Thus, when images are not
offset by whole integer pixels, the bias is manifested by
an inflated standard error. In our previous work
\citep[see Fig. 3 in][]{casettiwfpc2}
we have used a diagnostic tool somewhat different from a simple histogram
to explore this bias; namely, a differential curve based on the
distribution of pixel fractions compared to that of a uniform sample. Here, we
are able to present
a more direct test based on the many repeated images within the
Gilliland data set. While in \citet{casettiwfpc2} we explored
various {\bf classic} centering algorithms, and established that
the hst1pass performed best, here we show that the pixel-phase
bias is still present.  Its presence is most likley due to a slight mismatch
between the library ePSF and that of the actual
observations under consideration.
Note that we have also built similar curves using positions corrected for the 34th-row
and optical distortions; their shapes are nearly-identical
to those in Fig. \ref{fig:stderror-hst1pass}. 
Thus, we conclude the dominant cause of the striking coherence of these curves is the
presence of pixel-phase bias, and it is this aspect that we focus on in this work.
For reference, we show in Figure \ref{fig:q-vs-mag} the run of
the quality-of-fit parameter q as a function of instrumental magnitude for
the PC and one WF chip (WF3). Objects that participate in calculating the standard errors
shown in Fig. \ref{fig:stderror-hst1pass} are, by far, predominantly those
with good q values. Thus, the variation of standard errors with offsets is not attributable to
the inclusion of sources with poor-quality fits (such as saturated objects,
or too-faint objects).

\section{Centering Star Images with a Deep Learning Algorithm \label{sec:dla}}
\subsection{Strategy \label{subsec:strategy}}
To our knowledge, Deep Learning (DL) has not previously been used
for the computation of star centers in CCD images for astrometric purposes.

We strive only to model the center position of a stellar image, {\bf not} its magnitude.
This is in order to simplify the approach at the outset and solely focus on
aspects that affect the stellar center. Certainly, magnitude determination
could be introduced into the process, but this is beyond the
scope of the present study.
Thus, in what follows, instrumental magnitudes
do not participate directly in the training
process and are only used to help restrict samples
to, for instance, well-measured objects.
The input for the training process
consists of the $(x,y)$ star center and the pixel intensity
map of that star. This map is basically a square raster cutout
around the star from the calibrated fits file of a given exposure.
We have tested various raster sizes and eventually
decided to use a $6 \times 6 $ pixel size as we explain in 
Sec. \ref{subsec:simres}. As far as the algorithm is concerned,
the stars' centers
are with respect to the bottom left corner of the raster. 

Initially we employ simulated images, as described in Sec. \ref{sec:sim}, where
the ground truth is exactly known.  After that, we perform tests with real images where,
in principle, the true positions are not known. Instead, we adopt a 
``ground truth'' based on a (sigma-clipped) average-position catalog
obtained from the 636 (645) images in the
F555W (F814W) filter as detailed in Sec. \ref{sec:real}.

Note that this approach does not make any assumption regarding the PSF shape,
but instead estimates the (x,y)
coordinates of the star center by measuring correlations in the pixel
intensity values within a square raster around each star.
Previous and related work done by our group, but which only includes
preliminary results from simulated data,
can be found in \citet{baena2022}. In that work,
we have explored two different image simulation codes, two
different PSF types and two different classic centering algorithms.
The major conclusion of that work is that the DL code provides an
astrometric precision superior to the classic centering algorithms.
More specifically, the DL code in \citet{baena2022}
shows an improvement of between
  20\% and 30\% compared to the hst1pass code.

\subsection{Deep Learning Model \label{subsec:specs}}

\begin{figure} 
\includegraphics[scale=0.30,angle=0]{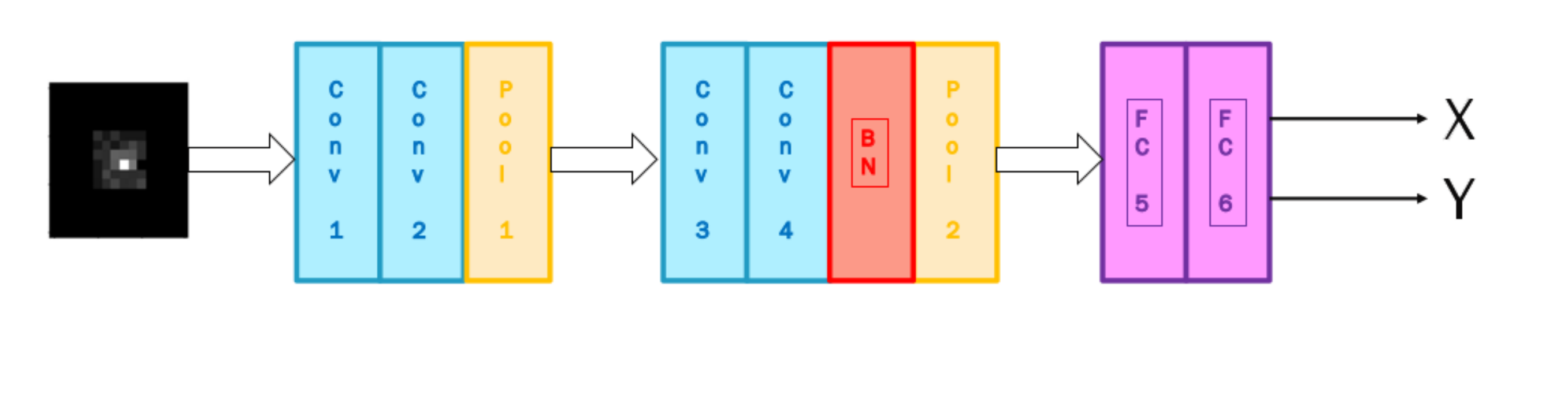}
\caption{VGG6 model architecture.}
\label{fig:vgg6}
\end{figure}

Our specific CNN model is a VGG \citep{simonyan14} with
six trainable layers, four of them convolutional layers plus two
fully-connected, and which end in two neuron outputs.
Each of these is to provide an estimate along the
x- and y-axis, respectively. One max-pool layer is inserted every two
convolutional ones, and all hidden layers are
equipped with ReLu non-linearity except the last one
which is \textit{linear}, in accordance with the regression
nature of this problem. We found that inserting a
batch-normalization layer after the fourth convolutional
one yields better results and helps
stabilize the convergence process in our specific problem.
The final architecture is illustrated schematically in Figure \ref{fig:vgg6}.

Due to the finite dimensions of the original
cutout image of $6 \times 6$ pixels\footnote{Various sizes of the
raster cutout are explored in Sec. \ref{subsec:simres}},
the model is limited in depth. Therefore, we must limit the total number of layers,
which under other circumstances would have allowed us to increase
the level of detail that can be analyzed or
the number of features that can be extracted
from the data. In our case, we have instead opted to modify the
number of trainable parameters
by increasing the number of kernels at each layer.
This allows us to check how the model behaves with
respect to the number of parameters,
in terms of overfitting. This is critical at this
stage since our assumptions are that 1) stars are isolated within their cutout
region, with no nearby source contamination, and
2) the PSF does not vary across the chip.

To this end, two different VGG models are trained, the first one
with 34K parameters (hereafter called V1),
and a second one with 214K
parameters (called V2). We found that V2
gives better results but with some overfitting
effect during training.
The $6 \times 6 $ pixel rasters are also
zero padded to guarantee a
minimum number of layers as the
model increases in depth.

The input data set of stars is divided into subgroups of $70:10:20\%$
for training, validation, and final testing, respectively.
All input cutout images' intensities are normalized to a sum of one,
regardless of the noise level or whether the star is saturated.
The minimization of the cost function
(Mean Absolute Error) benefited from
low learning rates ($10^{-5}$),
while epochs had to be increased
up to 1000.
The model was designed in Keras/TF
\citep{chollet2018} and optimized with
ADAM using default values.

The specific properties of the VGG6 models used here are
listed in Table \ref{tab:vgg6-specs}.


\begin{deluxetable}{lll}
\tablecaption{VGG6 models \label{tab:vgg6-specs}}
\tablewidth{0pt}
\tablehead{
    \colhead{Layer (type)} &
    \colhead{Output shape} &
    \colhead{$\#$ of param.} 
}
\startdata
\multicolumn{3}{c}{V1 model - 34K parameters} \\
\hline
conv1 (Conv2D) & (None, 14, 14, 16) & 160 \\
conv2 (Conv2D) & (None, 12, 12, 16) & 2320 \\
max-pool2d (MaxPooling2D) & (None, 6, 6, 16) & 0 \\
conv3 (Conv2D) & (None, 4, 4, 32) & 4640 \\
conv4 (Conv2D) & (None, 2, 2, 32) & 9248 \\
bn2 (BatchNormalization) & (None, 2, 2, 32) & 128 \\
max-pool2d-1 (MaxPooling2D) & (None, 1, 1, 32) & 0 \\
flatten (Flatten) & (None, 32) & 0 \\
fc-1 (Dense) & (None, 512) & 16896 \\
fc-out (Dense) & (None, 2) & 1026 \\
\hline
\multicolumn{3}{l}{Total param.: 34418; trainable param.: 34354} \\
\hline
\hline
\multicolumn{3}{c}{V2 model - 214K parameters} \\
\hline
conv1 (Conv2D) & (None, 16, 16, 32) & 832 \\
conv2 (Conv2D) & (None, 12, 12, 32) & 25632 \\
max-pool2d (MaxPooling2D) & (None, 8, 8, 32) & 0 \\
conv3 (Conv2D) & (None, 6, 6, 64) & 18496 \\
conv4 (Conv2D) & (None, 4, 4, 64) & 36928 \\
bn2 (BatchNormalization) & (None, 4, 4, 64) & 256 \\
max-pool2d-3 (MaxPooling2D) & (None, 2, 2, 64) & 0 \\
flatten-1 (Flatten) & (None, 256) & 0 \\
fc-1 (Dense) & (None, 512) & 131584 \\
fc-out (Dense) & (None, 2) & 1026 \\
\hline
\multicolumn{3}{l}{Total param.: 214754; trainable param.: 214626}
\enddata
\end{deluxetable}

\section{Application to Simulated Data \label{sec:sim}}
\subsection{Simulation Code \label{subsec:scode}}
As a precursor to exploring the effectiveness of the VGG
centering on real WFPC2 images, an in-house
image simulation code was written to
construct star-field images that specifically
emulate processed WFPC2
exposures.  Given an ePSF profile --
such as those used in the 
standard hst1pass source detection and centering
software -- and an input 
list of star positions and magnitudes, simulated
image files are produced for 
each of the four WFPC2 chips.  The simulation code
includes the effects of 
gain, exposure time, field-dependent sensitivity
and vignetting (via actual 
WFPC2 flatfield-correction files), A/D saturation,
column bleeding, charge 
transfer inefficiency charge shifts (CTI), sky background,
cosmic rays, and 
bias and dark current.  Poisson noise is added for
the appropriate sources of 
such, as well as Gaussian readout noise.
Finally, the images are given bias, 
dark-current, and flatfield corrections
yielding the equivalent of processed 
WFPC2 images.

Note that actual WFPC2 images exhibit a field dependent ePSF
requiring interpolation within a spatial grid 
of ePSFs, in practice.  For the trials presented
here we restrict ourselves to a uniform ePSF for
each chip.  In effect, all of the simulated star images are
as if they were found within the central portion of a real WFPC2 chip.
Also, CTI and cosmic rays
have been "turned off" in the present 
simulations. For these simulation tests we use the hst1pass
library central ePSF for
filter F555W to construct constant-ePSF
images of the PC and WF3 chips in this filter.

Our simulation code does {\bf not} include the 34th-row
systematic error or
the astrometric effects of optical distortion, as we are
concerned here only with the PSF modeling.

\subsection{Simulation Input \label{subsec:simin}}
To generate a WFPC2-type exposure we start with an input catalog
of positions and magnitudes. We note that throughout this discussion
magnitudes are instrumental (uncalibrated) magnitudes. 
We begin with the filter F555W test, for which the input catalog is created as follows.
One of the F555W 160-sec exposures is processed with hst1pass ---
specifically the PC chip --- 
to produce a list of detections with input $(x,y)$ positions
and instrumental magnitudes. We include all objects with quality
parameter q = 0.0 to 1.0. Note that objects with q = 0.0 are
detected but not precisely centered by hst1pass as they are deemed saturated;
hst1pass assigns such objects a fractional pixel-position of 0.
This does not matter since we discard the fractional part of the $(x,y)$ positions of
{\bf all} objects in the list,
and assign randomly chosen pixel fractions. This ensures
there is no fractional-pixel bias in our input catalog.
This input list is then used to generate a
PC and a WF3 image with the simulation code described in
Sec. \ref{subsec:scode}. The magnitude distribution of this input
list is the natural magnitude distribution of stars in cluster
NGC 104, and as such most objects are toward fainter magnitudes.
Thus, the training process will be dominated by stars with
rather faint magnitudes. We will refer to this
sample as ``natural''.

However, we also want to better explore the performance of the
DL algorithm at the bright end. Accordingly, we have
created a second input catalog with an artificial magnitude
distribution, such that bright stars are overly represented.
We refer to this sample as ``bright-enhanced''.

\subsection{Simulation Results \label{subsec:simres}}

In Table \ref{tab:input-numbers} we show the total
number of input objects, (N$_{tot}$), that went into the DL modeling process.
The training+validation sample uses $80\%$ of these objects,
while the testing is done on the remaining $20\%$.
The total number of testing objects, (N$_{test}$), is shown in column 3
of Tab. \ref{tab:input-numbers}. In column 4, (N$_{eval}$),
we indicate the average number of objects used to
evaluate the standard deviation of differences in position
between the input (true) values and the DL-determined ones.
This latter number is derived from the N$_{test}$ sample,
but trimmed within a fixed magnitude range --- to ensure we are dealing
with well-measured stars --- and further trimmed by
eliminating $3\sigma$ position-difference outliers in an iterative procedure.
The magnitude range of N$_{eval}$ for the natural-magnitude
distribution sample is thus limited to the subset of stars with
magnitudes between 
15.0 and 18.0, and for the bright-enhanced magnitude
distribution sample to between 13.0 and 15.0.

\begin{deluxetable}{lrrr}
  \tablecaption{Simulated-data sample statistics}
  \label{tab:input-numbers}
\tablewidth{0pt}
\tablehead{
    \colhead{Sample} &
    \colhead{N$_{tot}$} &
    \colhead{N$_{test}$} &
    \colhead{N$_{eval}$} 
}
\startdata
Natural  &  4648  & 930 & $\sim 600$   \\
Bright-enhanced & 8493 & 1699 & $\sim 530$ \\ 
\enddata
\end{deluxetable}

\begin{figure} 
\includegraphics[scale=0.45,angle=0]{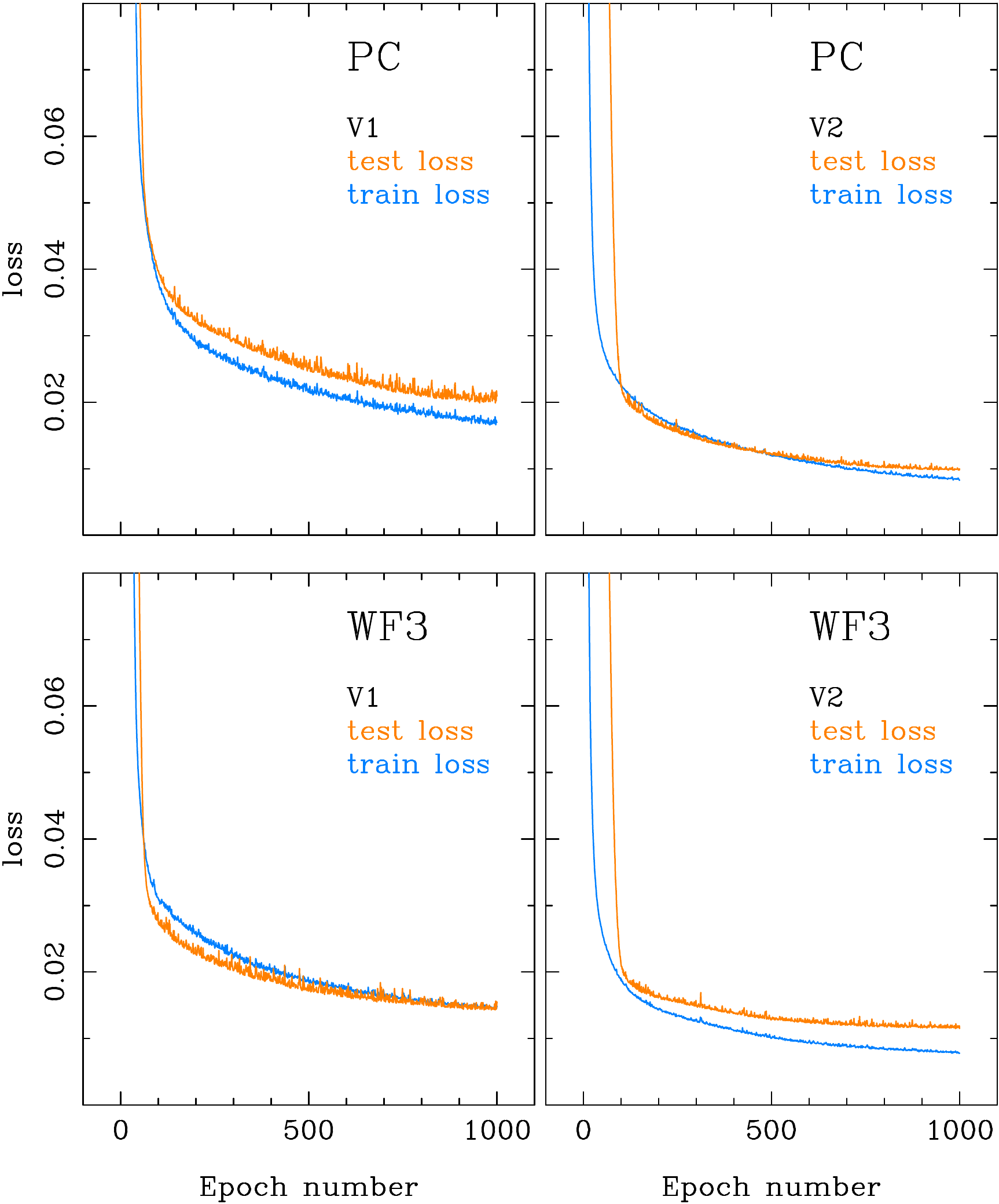}
\caption{Simulated data for the natural-magnitude distribution sample:
  the loss function and its
  dependence on epoch
  for the $6\times6$-pixel raster run.
  Both PC and WF3 chips for models V1 and V2
  are shown.
  Recall that WF3 is more severely undersampled
  compared to the PC, by roughly a factor of 2.}
\label{fig:sim-loss}
\end{figure}

In Figure \ref{fig:sim-loss} we show the loss
function of the training and testing samples
as a function of model epoch.
This is run for the
natural-magnitude distribution sample and for 
a raster size of $6\times6$ pixels;
both V1 and V2 models are shown.
The plots indicate that model V2 achieves a
better loss trend than does model V1. 

\begin{figure*} 
\includegraphics[scale=0.65,angle=-90]{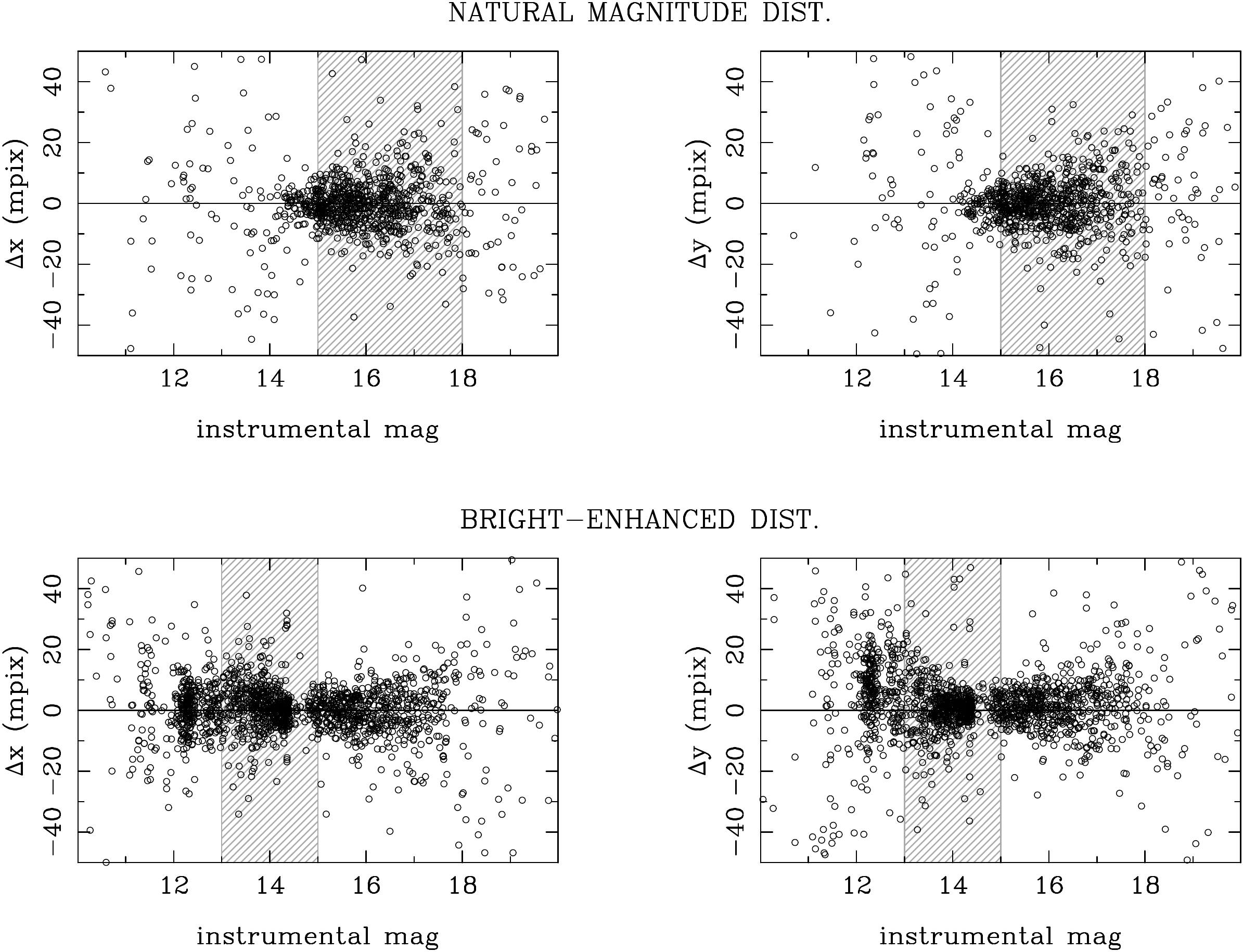}
\caption{Position differences (DL-derived minus true) as a
  function of magnitude for simulated data. This realization is
  for the PC, using a raster size of $6\times6$ pixels, and model V2.
  Both natural ({\it top}) and bright-enhanced ({\it bottom})
  magnitude distribution samples are
  shown. The hashed areas highlight the restricted magnitude range
  over which standard deviations are calculated.}
\label{fig:sim-pos-diff}
\end{figure*}

Next, we plot position differences as a function of
magnitude for both the natural and bright-enhanced samples
in Figure \ref{fig:sim-pos-diff}.
The realization shown is for the PC with raster size of $6\times6$ pixels
and for version V2 of the DL model (see \ref{subsec:specs}).
In the bright-enhanced sample, the larger scatter in $\Delta y$
compared to that in $\Delta x$ at magnitudes $\le 13.5$ is due to
saturation and bleeding effects along the read-out direction of the
chip. Nevertheless, the bright end still has a tight
position-difference distribution for the sample that includes
many more bright stars in the training process. This
indicates that the bright stars can be reliably DL-modeled,
if special attention is given them.

In what follows, we will use the standard deviation of
the position differences as a measure of the error in
each model realization. We use values obtained from
samples within the restricted magnitude ranges listed above,
and with $3-\sigma$ outlier clipping. We explore these
errors as a function of the raster size cutout around each
object. Simulations for both the PC and WF3 are tested,
and for both V1 and V2 DL models. Again, both the natural and
bright-enhanced samples are studied. 

\begin{figure*} 
\includegraphics[scale=0.65,angle=-90]{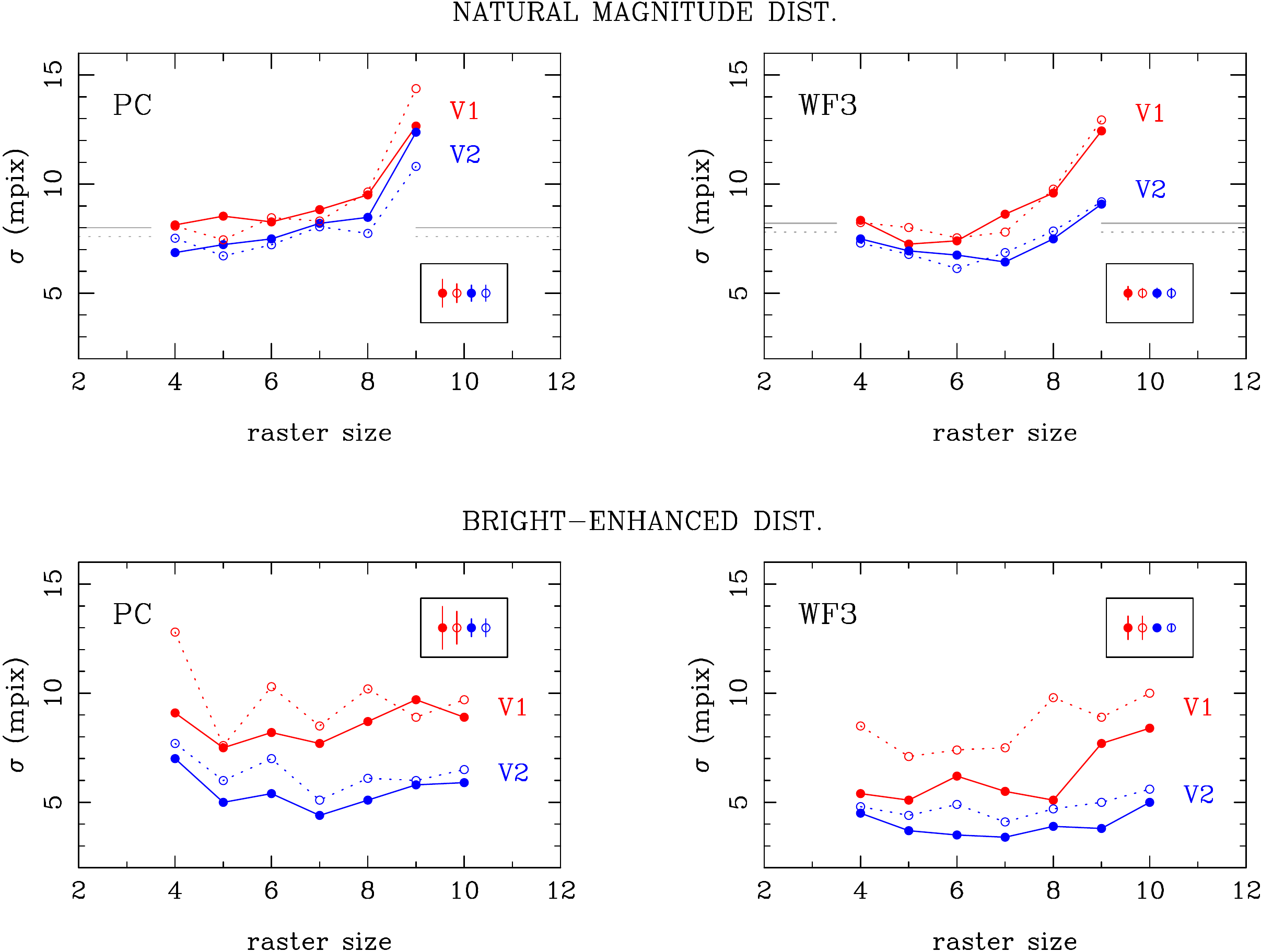}
\caption{Standard deviation of position differences as a function of
  raster size for simulated PC and WF3 images. Continuous lines
  correspond to the $x$ coordinate, and dashed lines to the $y$ coordinate,
  The two versions of the DL model are shown.
  Also, both natural and bright-enhanced magnitude
  distribution samples are plotted, as labeled.
  For the natural sample we also indicate with
  horizontal solid and dashed lines the standard deviations in each
  coordinate obtained from hst1pass-based positions.
  For the bright-enhanced sample there are no such values, since hst1pass is
  not designed to center these stars.
  Typical error bars obtained for the $6\times6$-pixel raster
  experiment are given in the legend box.}
\label{fig:sim-raster}
\end{figure*}

The
standard deviations of position differences are shown in
Figure \ref{fig:sim-raster}. Error bars were estimated
for the $6\times6$-raster experiment by repeating the experiment
five times, i.e., the uncertainty estimates are based on the
scatter produced by the
randomness of the selection of the $20\%$ test sample in the DL code.
For comparison, we also indicate the value of the standard deviation 
calculated from hst1pass-derived positions obtained
on these same simulated images, within the restricted central 1/9 area
of the chip. We use raw hst1pass positions, i.e., uncorrected for
34-row and optical distortion, since the simulations do
not incorporate these effects. The generic hst1pass code
was applied only to the natural magnitude distribution
sample, as it is not optimized for bright objects.

Fig. \ref{fig:sim-raster} indicates that version V2 of the DL model
gives better results than does V1; this is in agreement with the
indication by the loss function shown in Fig. \ref{fig:sim-loss}.
Small raster-size cutouts tend to give better results than do large
ones, for the natural magnitude sample. For WF3, there appears
to be a minimum at a raster size of around 6 pixels, while for the PC
this value is also near minimum. Therefore, in our
exploration of real data, we will adopt a fixed $6\times6$ raster size
for both the PC and the WF chips. 

The hst1pass standard deviations are in
reasonable agreement with the low portion of the DL values, at raster sizes between
4 and 8 pixels.  This is not surprising, since
the simulated images were built with an ePSF that is exactly that from the hst1pass
standard library. In fact, for the WF3 chip, the DL V2 results are even
somewhat better than those of hst1pass. Finally,
for the bright-enhanced sample, it is evident that excellent
results can be obtained with the DL methodology, given
adequate input training data.

\section{Application to Real Data \label{sec:real}}
\subsection{Input and DL model \label{subsec:realin}}
For real WFPC2 images it is clear that we do not readily
have ``true'' positions in order to train the
DL technique. 
Is it possible to form a catalog of positions that are sufficiently
precise to serve this purpose?
The NGC 104 Gilliland
data set offers such a possibility; with over
600 exposures taken at various small 
offsets (up to 2 PC pixels, see Fig. \ref{fig:offsets})
an average catalog can be built
and used as ``true'' positions for training.
We construct this average catalog by first transforming the positions from each exposure
into those of the first exposure thus placing all exposures on the
system of the first one. The positions to be used are those from the hst1pass runs.
Only objects with quality parameter q = 0.0001 to 1.0 are used;
thus bright stars, with inferior hst1pass centers, do not participate in this process.
The polynomial transformation between reference and target exposure
includes up to 3rd order terms.
Once all are on the same system, the repeated measures for the same star are
combined to form a mean position, after iteratively removing any measures
more than $3\sigma$ from the mean.
The resulting average catalog's positions can then be transformed back
to the system of each individual exposure, providing what will serve as a ``true''
position to correspond to each star's cutout raster.

To avoid having to consider effects from the variable PSF across each WFPC2 detector, 
only stars in the central part of each chip are used, specifically, the central third of both
the $x$ and $y$ coordinate range.
We further refine the input list by 
discarding all objects that have a neighbor within
5 pixels, such that random blends
and crowdiness do not
confuse the training process.

Note that the large number of exposures,
and the distribution of fractional-pixel offsets among them,
will serve to effectively average out any pixel-phase bias present in the hst1pass positions,
as well improving the precision well beyond that of any single exposure.
It is this that allows us to use the hst1pass-based positions to form a sufficiently
precise and systematic-free set of ``true'' positions to do DL modeling.

In practice, for each exposure we extract rasters of
$6\times6$ pixels around each object in the input list.
The DL model can then be trained and verified using the average catalog positions ---
always in the system of the corresponding exposure --- along with the cutout rasters.
Although confined to the central area of each chip, the 636 exposures in filter F555W
provide an ample amount of input data to construct a DL model.
A separate run/model is made for each of the four WFPC2 chips.
The model architecture is the same as that used for the simulated data,
i.e., the two versions of VGG6 described in Sec.
\ref{subsec:specs}.

In a similar manner, separate DL models can also be constructed
based on the F814W data set, which has
654 exposures. The number of useful stars per exposure and chip in the input lists
is between 350 and 400 objects for the F555W data set,
and between 430 and 480 for the F814W data set.
The total numbers of objects that enter into the model
building for the entire set of over 600 exposures are
listed in Table \ref{tab:real-lists} for each chip and filter.
These numbers are a factor of between $\sim 50$ and 70 larger than
the number of input objects in the simulation experiments
(see Tab. \ref{tab:input-numbers}).
\begin{deluxetable}{lccccc}
  \tablecaption{Number of input objects per chip and filter
    \label{tab:real-lists}}
\tablewidth{0pt}
\tablehead{
    \colhead{Filter} &
    \colhead{N$_{exp}$} &
    \colhead{N$_{PC}$} &
    \colhead{N$_{WF2}$} &
    \colhead{N$_{WF3}$} & 
    \colhead{N$_{WF4}$}  
}
\startdata
F555W  & 636 & 235912  & 231391 & 253207 & 224953   \\
F814W  & 654 & 289430  & 297036 & 310076 & 282431   \\ 
\enddata
\end{deluxetable}

\begin{figure} 
\includegraphics[scale=0.47,angle=0]{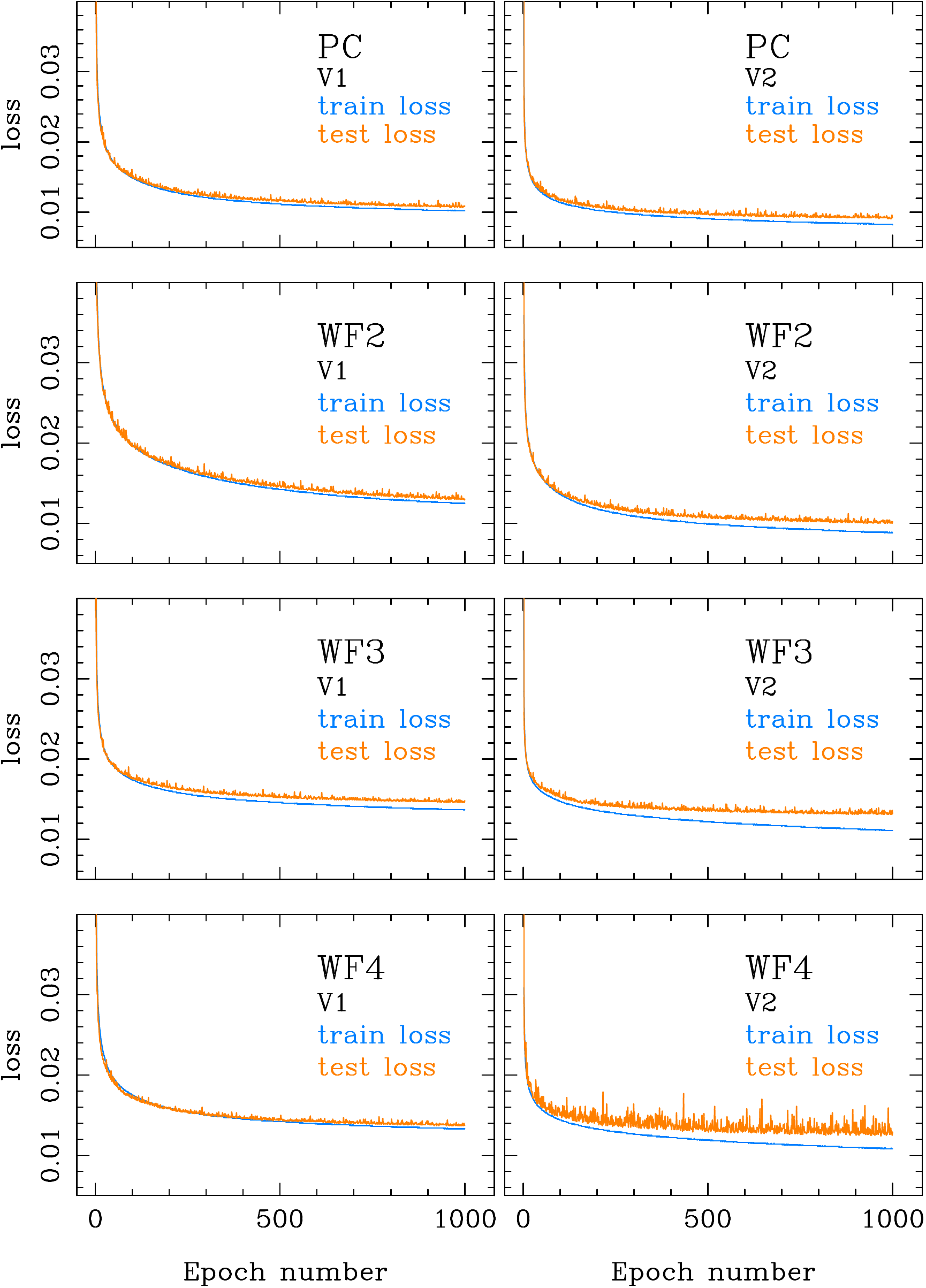}
\caption{Real F555W data: the loss-function trend with
  epoch for the four chips and
  two versions of the DL model.}
\label{fig:real-loss-f555w}
\end{figure}

\begin{figure} 
\includegraphics[scale=0.47,angle=0]{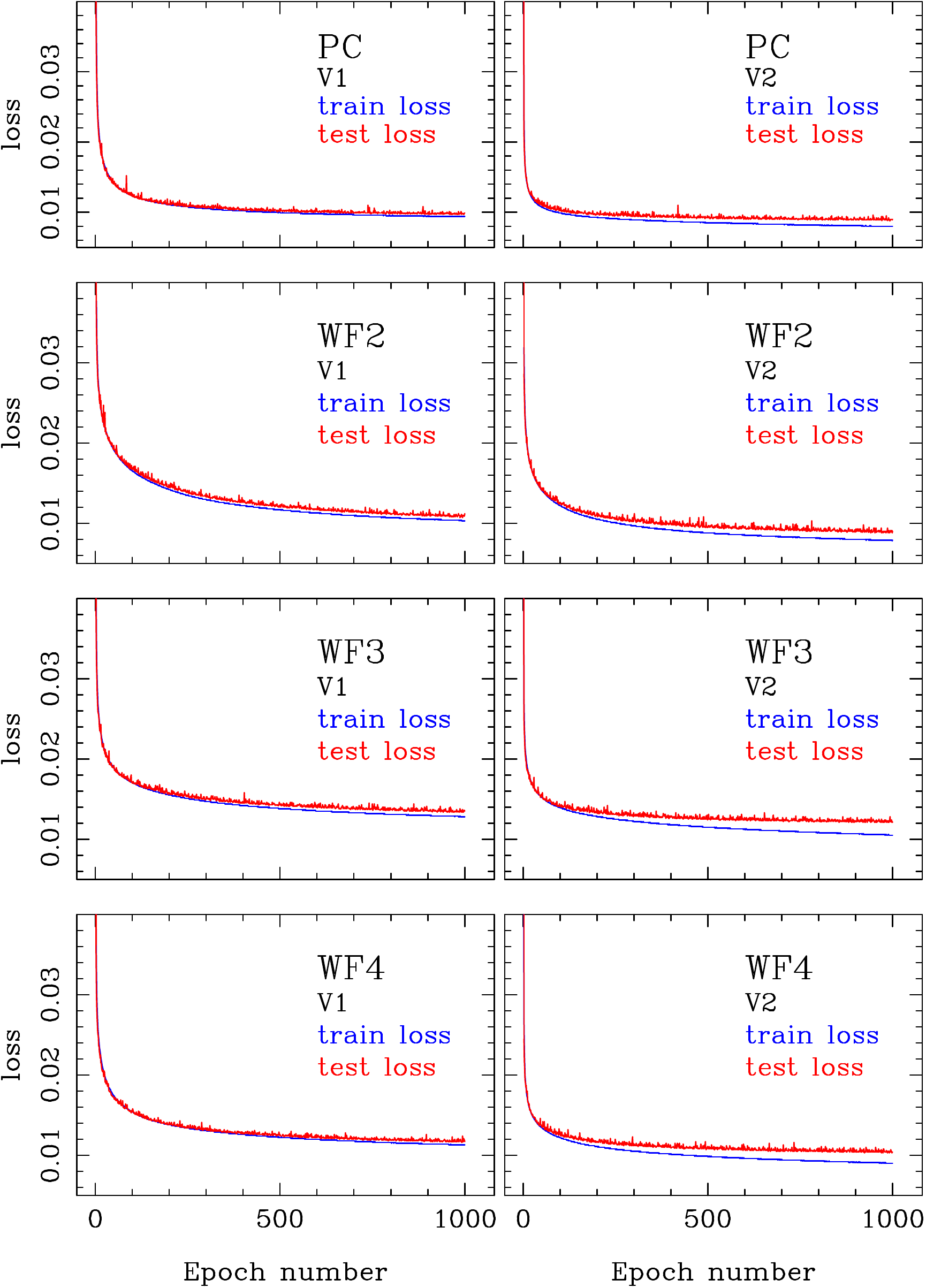}
\caption{Real F814W data: the loss function trend with
  epoch for the four chips and
  two versions of the DL model.}
\label{fig:real-loss-f814w}
\end{figure}

Figure \ref{fig:real-loss-f555w} shows the resulting loss-function trends with epoch
for the F555W data.
Each panel represents a different chip and version of the DL model, as labeled.
Figure \ref{fig:real-loss-f814w} shows the equivalent information for the DL modeling
of the F814W data.
The second version of the model appears to
perform better than the first version, especially for the PC detector.

\subsection{Real-Data Results \label{subsec:realres}}
The models derived in Sec. \ref{subsec:realin} can be applied to all (hst1pass) detections
in all exposures of the real data sets upon which they were built,
namely the NGC 104 Gilliland sets in F555W and F814W, provided one also adheres to
the central-area limits imposed by the presumption of non-variance of the PSF.
It would be interesting to see if these same models, derived from NGC 104 data, can
be applied to other WFPC2 data sets that have sufficient numbers of offset exposures and
stars per exposure to allow evaluation of the astometric accuracy of the
resulting star centers.

To this end, a search was made of the WFPC2 archive of observations to identify
any such data sets.
Specifically, we require data sets of targets sufficiently rich in measurable stars and 
containing a number of repeated exposures with small offsets. The repeated, offset exposures
will allow for the detection of any pixel-phase bias, while numerous
stars are needed to ensure a reliable transformation from
any given exposure into a chosen reference one, even while
considering only the central 1/9th of each chip.
Not surprisingly, there are very few such data sets in the archive,
and these all are in dense areas of not-too-distant
globular clusters. In Table \ref{tab:data-sets}
we summarize the properties of candidate data sets
in each of the two filters. The NGC 104 data sets used to build the models
are listed first, in bold letters, while the subsequent targets
are listed by observation date.
\begin{deluxetable}{lrrc}
  \tablecaption{Data sets used to test the DL models
    \label{tab:data-sets}}
\tablewidth{0pt}
\tablehead{
    \colhead{Target field} &
    \colhead{N$_{exp}$} &
    \colhead{Exp. time} &
    \colhead{Epoch} 
}
\startdata
\multicolumn{4}{c}{F555W} \\
\hline
\bf{NGC 104}  & {\bf 636} & {\bf 160}  & {\bf 1999.5}    \\
NGC 6752 - PC & 118 &  26  & 1994.6    \\ 
NGC 6441      &  36 & 160  & 2007.3    \\ 
NGC 6341      &  28 & 100  & 2008.1    \\
\hline
\multicolumn{4}{c}{F814W} \\
\hline
\bf{NGC 104}  & {\bf 654} & {\bf 160}  & {\bf 1999.5}    \\
NGC 6752 - PC & 109 &  50  & 1994.6    \\
NGC 6656      & 170 & 260  & 1999.1-2000.1    \\ 
NGC 6205      &  25 & 140  & 1999.8    \\ 
NGC 5139      &  24 &  80  & 2008.1    \\ 
\enddata
\end{deluxetable}

The appropriate DL models, by filter and chip, were applied to each of these sets.
Star detection and cutout rasters were made using hst1pass star centers as rough,
preliminary positions and confining the test samples to the central area of each chip.
Polynomial transformations of star positions from each exposure into those of a selected 
reference exposure were then made in order to evaluate the standard errors of 
the DL-model positions.

We present the results for our test data sets in plots similar to those in Fig.
\ref{fig:stderror-hst1pass}. However, in this case the
standard errors are plotted as a function of the pixel phase of the offset from the
reference exposure, instead of the full offset itself. 
In other words, this is the fractional part of the full offset, (shifting all offsets by
a constant, if need be, to ensure they are positive).   We designate this offset phase
$\phi$, and its value ranges from 0 to 1.

As such, in Fig. \ref{fig:stderror-hst1pass}, the
pixel-phase bias manifests itself as either a
``W'' or ''V''-shaped curve for the PC
and WF chips, respectively.
When plotting versus pixel-phase, these curves are essentially folded and the bias
shows itself as a curve with minimum standard error at $\phi$ = 0 and 1, and rising to 
a maximum at mid values of $\phi$.
In the absence of any pixel-phase bias,
the standard error curves should be flat as a function of
offset phase.

For each cluster and filter --- as specified in the
label above each plot --- we show these standard error curves
for positions obtained with hst1pass,
and with the VGG6 model.
For NGC 104 we show results for both versions of
the VGG6 model, while for the remaining clusters
we show only the VGG6 version V2 results,
as these were deemed the best.

\begin{figure} 
\includegraphics[scale=0.47,angle=0]{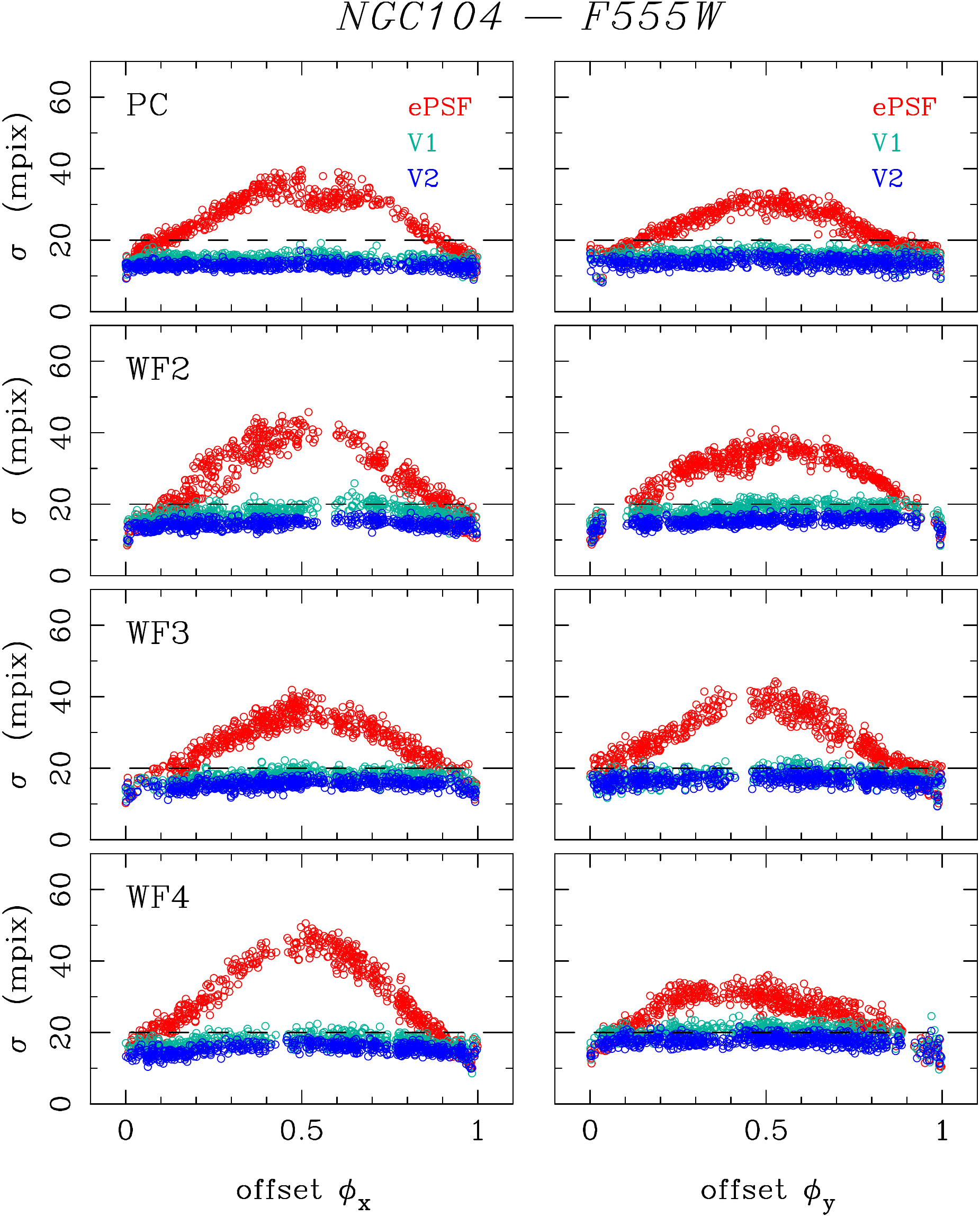}
\caption{Standard error of the transformation
  (of a target exposure into the reference exposure) as a
  function of offset phase. x and y- coordinate values are shown in the
  left and right panels, respectively, and each row represents
  a chip, as labeled. 
  A curve exhibiting larger standard errors near mid pixel-phase values indicates the 
  presence of pixel-phase bias in the positions.
  Three centering algorithms are presented and color-coded,
  as labeled: classic ePSF/hst1pass and DL VGG6 versions V1 and V2.
  Note the flat curves at $\sim 13- 15$ mpix obtained for the DL centers.
  These results are for exposures of NGC 104 in filter F555W.
}
\label{fig:f5-ngc104}
\end{figure}

The results for NGC 104 in F555W
presented in Figure \ref{fig:f5-ngc104}
show a vast improvement of the VGG6
positions over those obtained with hst1pass
in all four chips. Also version V2 of the VGG6 model
gives slightly better results than does version V1.
The DL positions show almost no pixel-phase bias.

Next, we present results for the other
data sets observed in filter F555W.
Keep in mind the DL models applied to these sets were based on the NGC 104 data,
while the target sets were generally taken at a different observation epoch.
Also, none of these sets have the number of repeats or 
excellent offset-phase $\phi$ coverage
of the NGC 104 data set, making the standard error plots noisier and less coherent. 
Finally, consideration of the observation
exposure times and inherent richness of some target clusters leads to a range in
overall quality of the standard-error plots.

\begin{figure} 
\includegraphics[scale=0.47,angle=0]{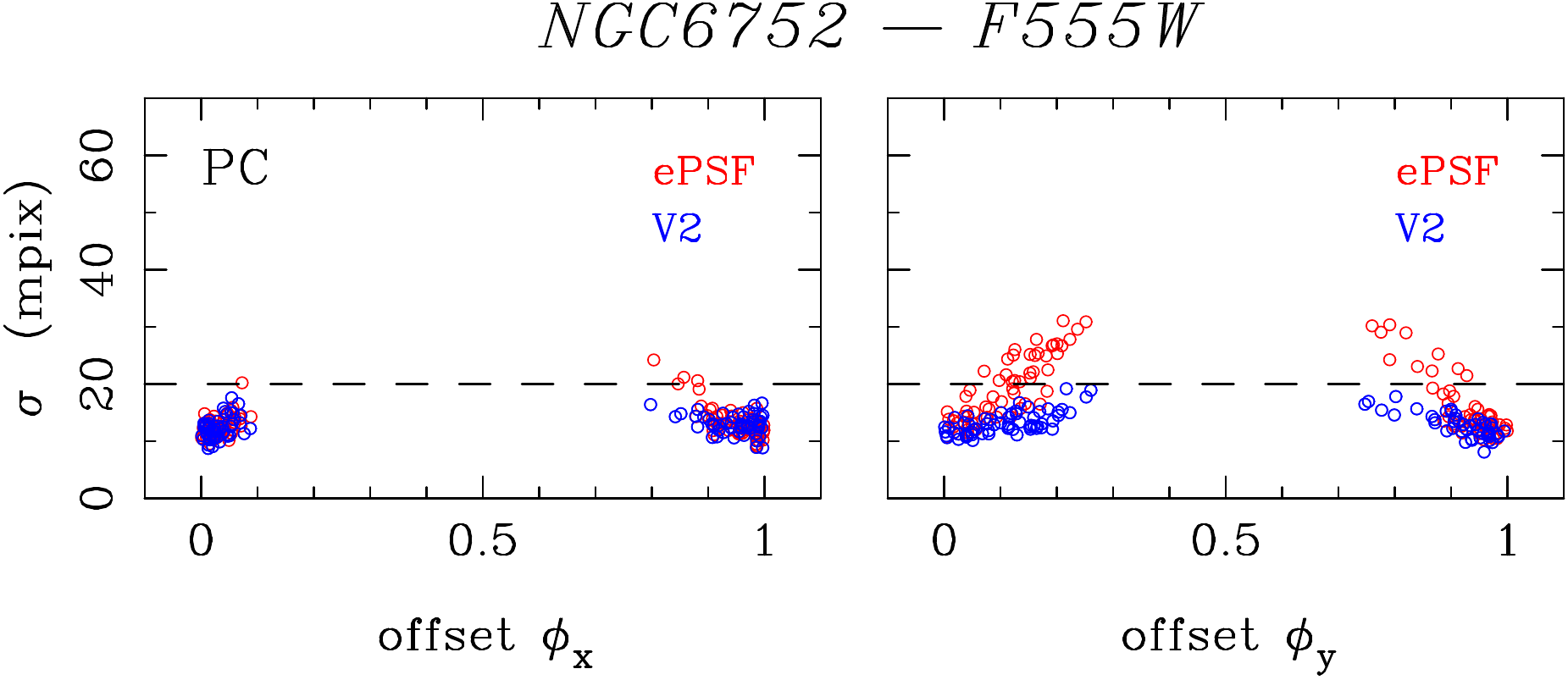}
\caption{As in Fig. \ref{fig:f5-ngc104}, only for NGC 6752 in F555W.}
\label{fig:f5-ngc6752}
\end{figure}

The data set for NGC 6752 had only PC
observations, in both filters.
This is a bit unfortunate, since this
data set is the only one taken at an earlier
epoch compared to the NGC 104 sets
(see Tab. \ref{tab:data-sets}).
Results shown in Figure \ref{fig:f5-ngc6752}
indicate that
the VGG6 model is an improvement for the F555W
data set.
However, the sampling is unfortunate as we miss the
mid portion of this figure where the presence of
pixel-phase bias shows itself best.

\begin{figure} 
\includegraphics[scale=0.47,angle=0]{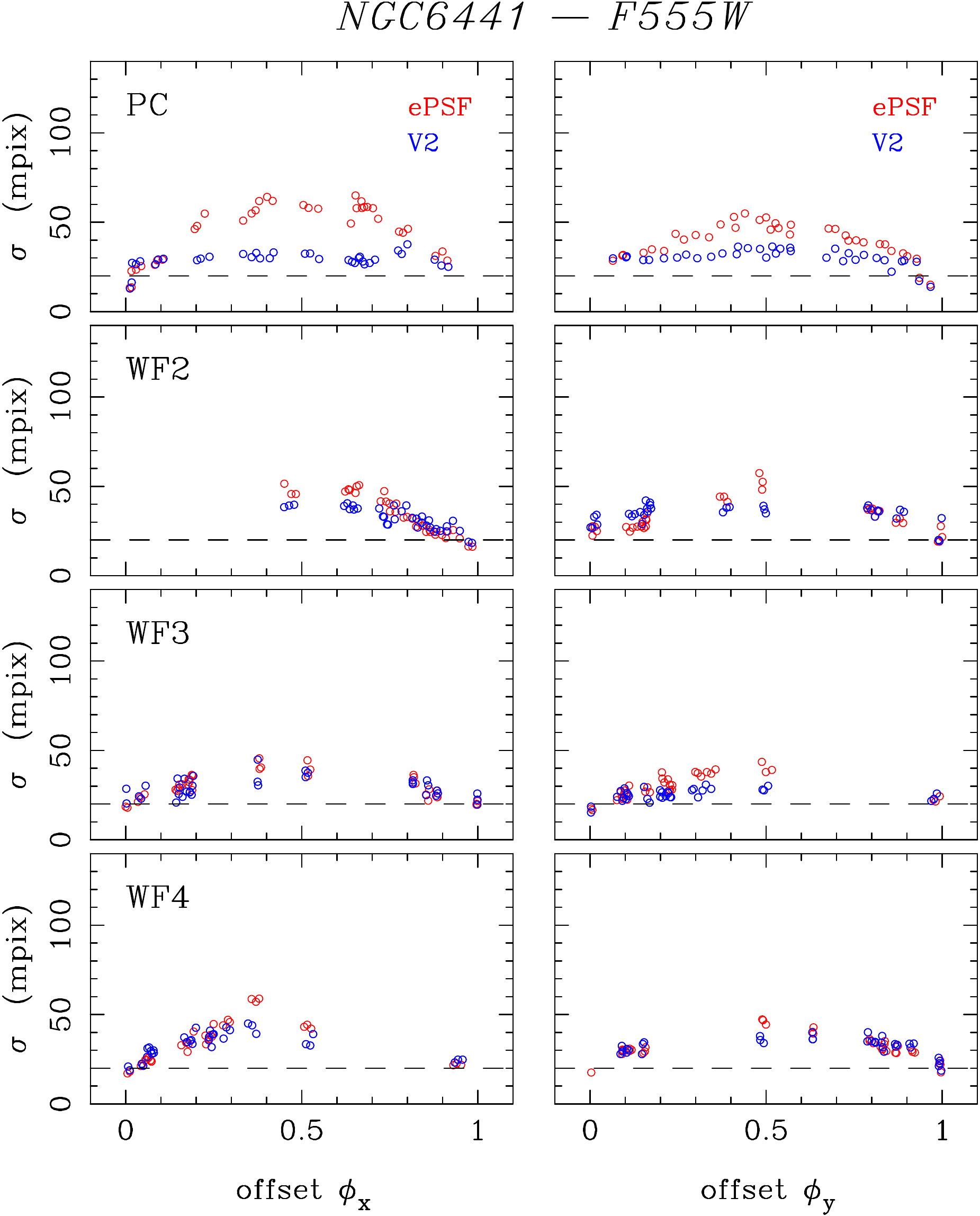}
\caption{As in Fig. \ref{fig:f5-ngc104}, only for NGC 6441 in F555W.}
\label{fig:f5-ngc6441}
\end{figure}

\begin{figure} 
\includegraphics[scale=0.47,angle=0]{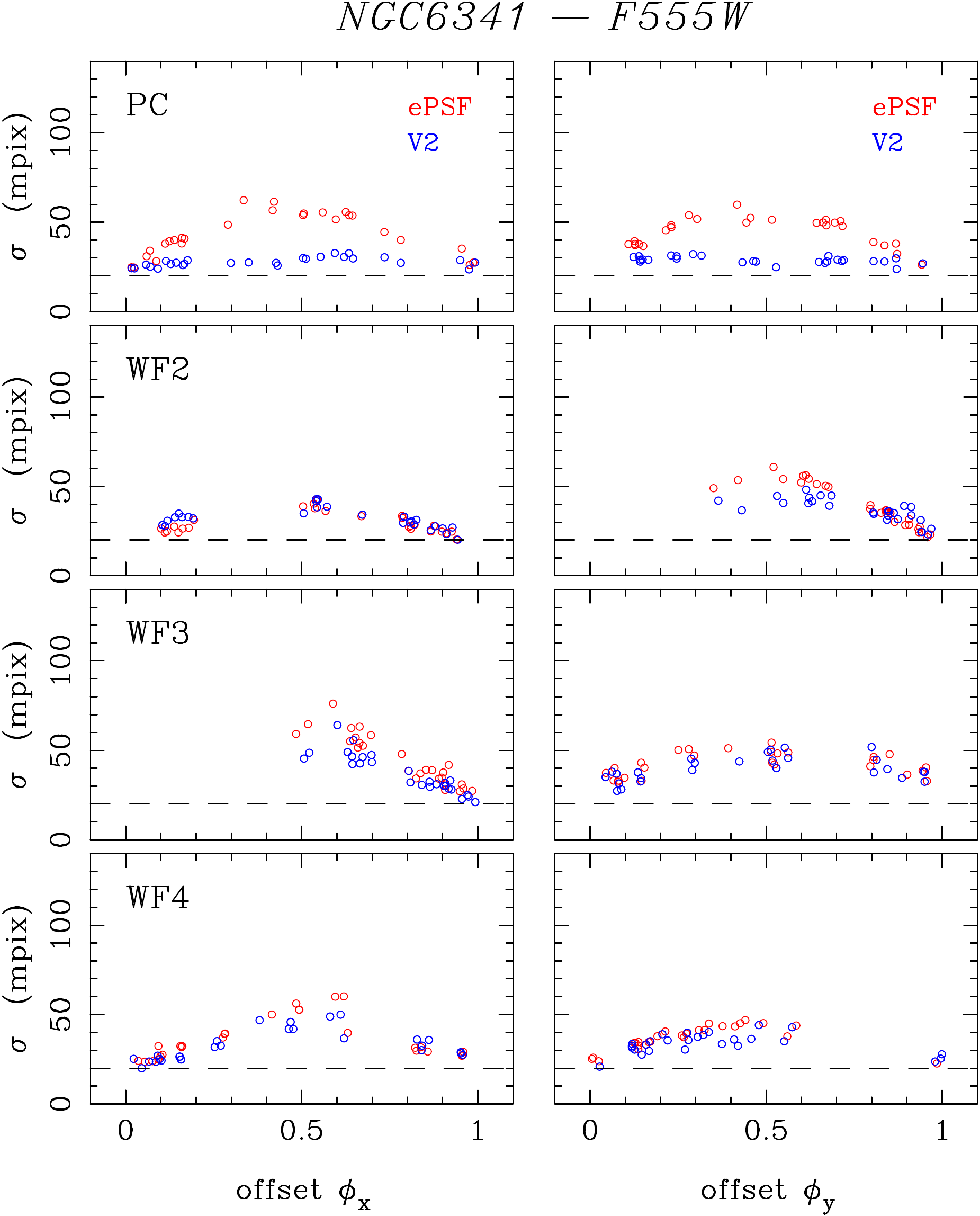}
\caption{As in Fig. \ref{fig:f5-ngc104}, only for NGC 6341 in F555W.}
\label{fig:f5-ngc6341}
\end{figure}

Results for NGC 6441 and NGC 6341 in filter
F555W are presented
in Figures \ref{fig:f5-ngc6441} and
\ref{fig:f5-ngc6341}. For the PC there is marked
improvement in the VGG6 positions compared to
the hst1pass ones. However, for the WF there is
only marginal improvement. Both these data sets
were taken at a much later epoch than the NGC 104
set. Presumably, the VGG6 PSF model based on
the NGC 104 data set is likely no longer
representative for the later sets, and this results in
the retention of some pixel-phase bias. This
is manifested more so in the WF chips since
these are the most undersampled.


\begin{figure} 
\includegraphics[scale=0.47,angle=0]{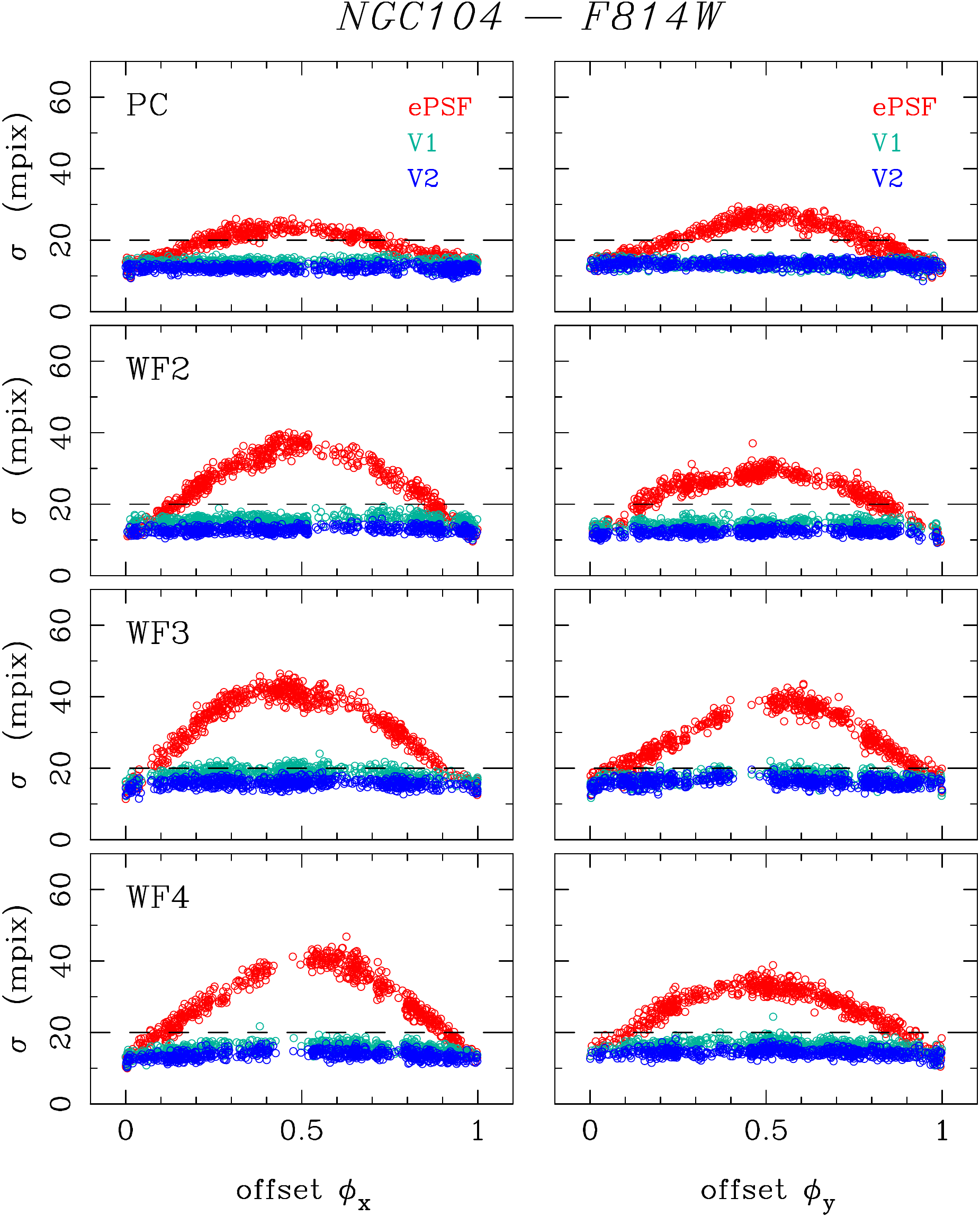}
\caption{As in Fig. \ref{fig:f5-ngc104}, only for NGC 104 in F814W.}
\label{fig:f8-ngc104}
\end{figure}

Turning our attention to the F814W data sets, 
Figure \ref{fig:f8-ngc104} shows the standard
errors of the various centering algorithms as applied to the NGC 104 
exposures in this filter.
The results are qualitatively similar to those of the F555W data set 
for this cluster, i.e., both V1 and V2 showing much less pixel-phase
bias compared to hst1pass.
Although, for the PC, the
improvement is not as dramatic in F814W as it was in F555W.
This is probably explained by the F814W images being better sampled
(especially in the PC) at these longer
wavelengths and, thus, it would be expected that any
pixel-phase bias present in the hst1pass centers would be less severe.

\begin{figure}
\includegraphics[scale=0.47,angle=0]{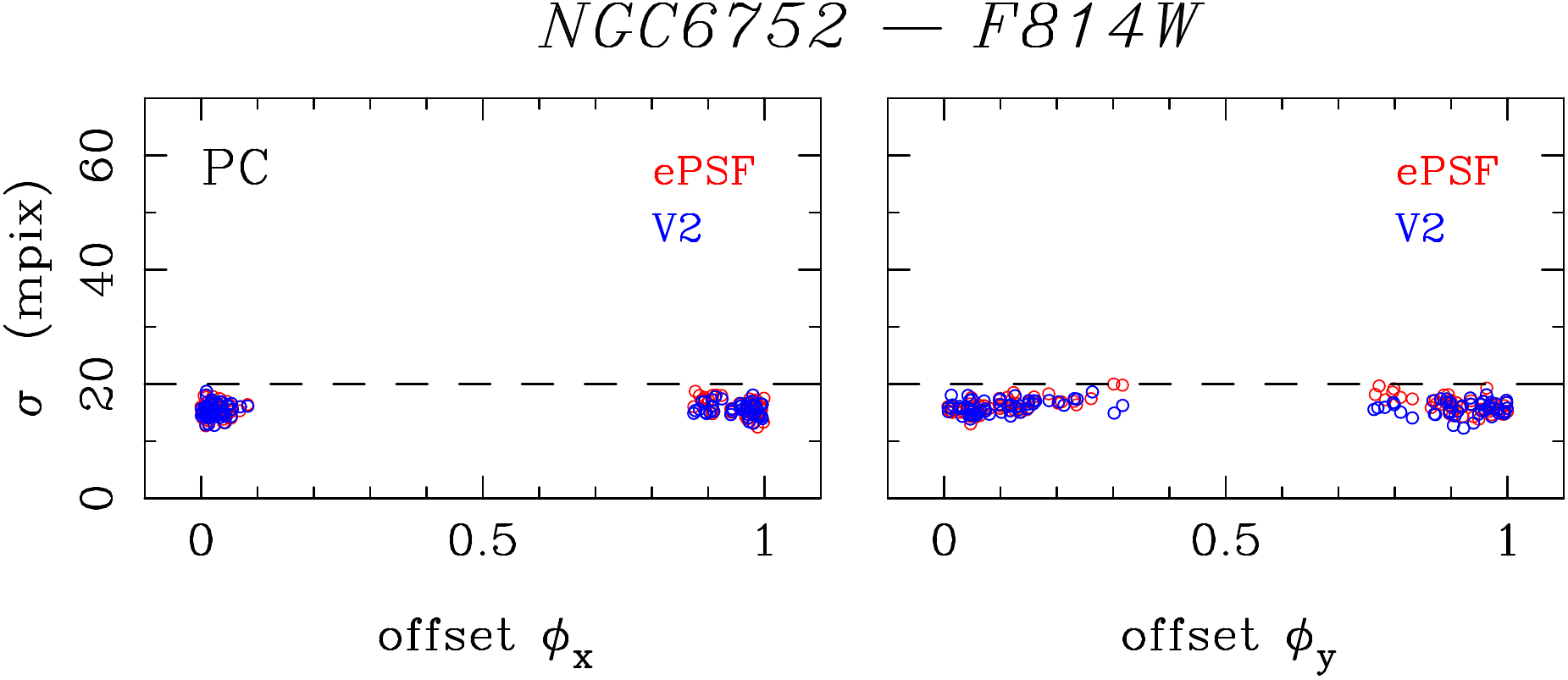}
\caption{As in Fig. \ref{fig:f5-ngc104}, only for NGC 6752 in F814W.}
\label{fig:f8-ngc6752}
\end{figure}

Next, we show the results for the PC-only F814W data set of NGC 6752 in
Figure \ref{fig:f8-ngc6752}.
Improvement in the standard errors of the DL positions over those from hst1pass is
barely perceptible, at best.
The limited pixel-phase coverage hinders any real conclusion here. 

\begin{figure} 
\includegraphics[scale=0.47,angle=0]{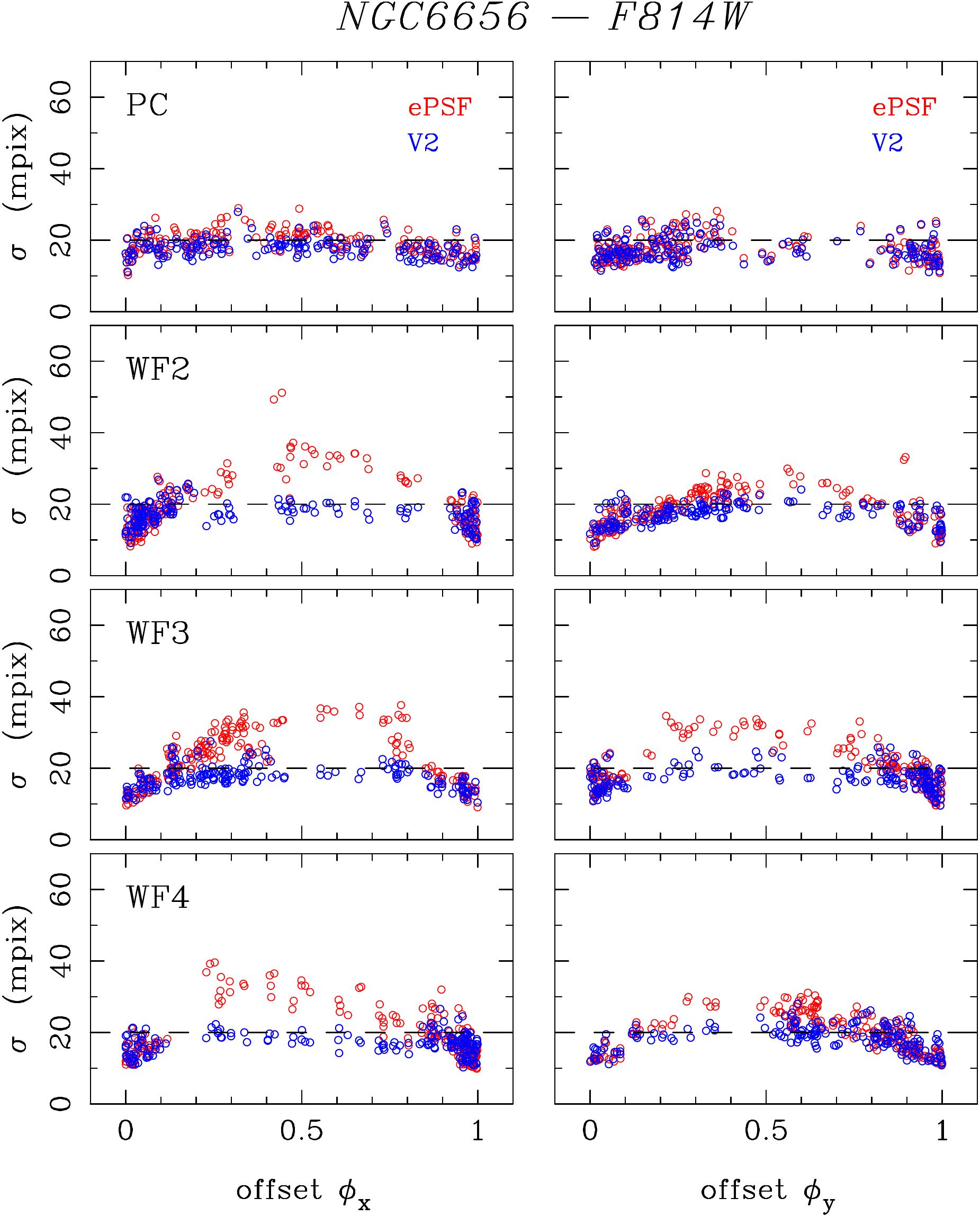}
\caption{As in Fig. \ref{fig:f5-ngc104}, only for NGC 6656 in F814W.}
\label{fig:f8-ngc6656}
\end{figure}

\begin{figure} 
\includegraphics[scale=0.47,angle=0]{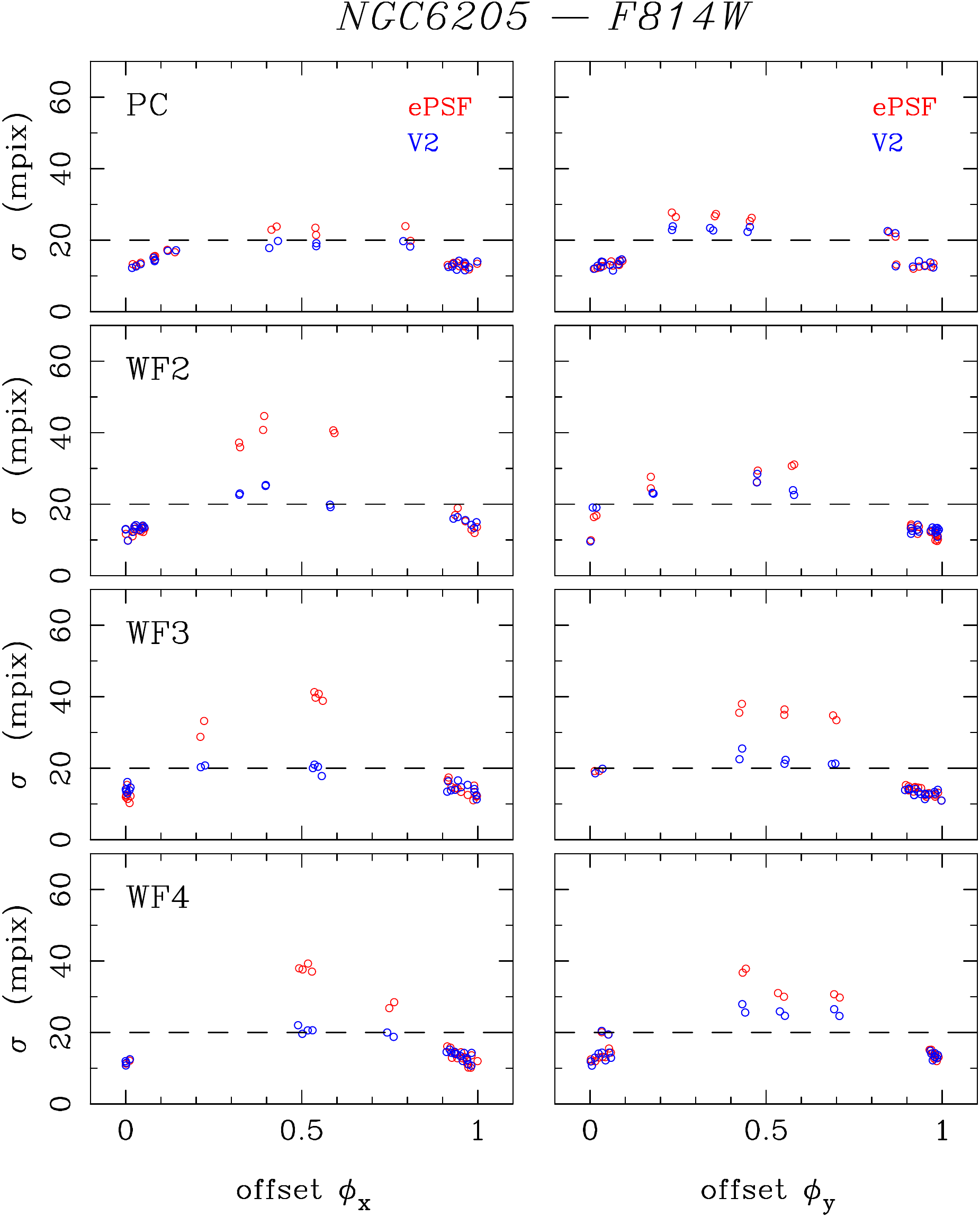}
\caption{As in Fig. \ref{fig:f5-ngc104}, only for NGC 6205 in F814W.}
\label{fig:f8-ngc6205}
\end{figure}

\begin{figure} 
\includegraphics[scale=0.47,angle=0]{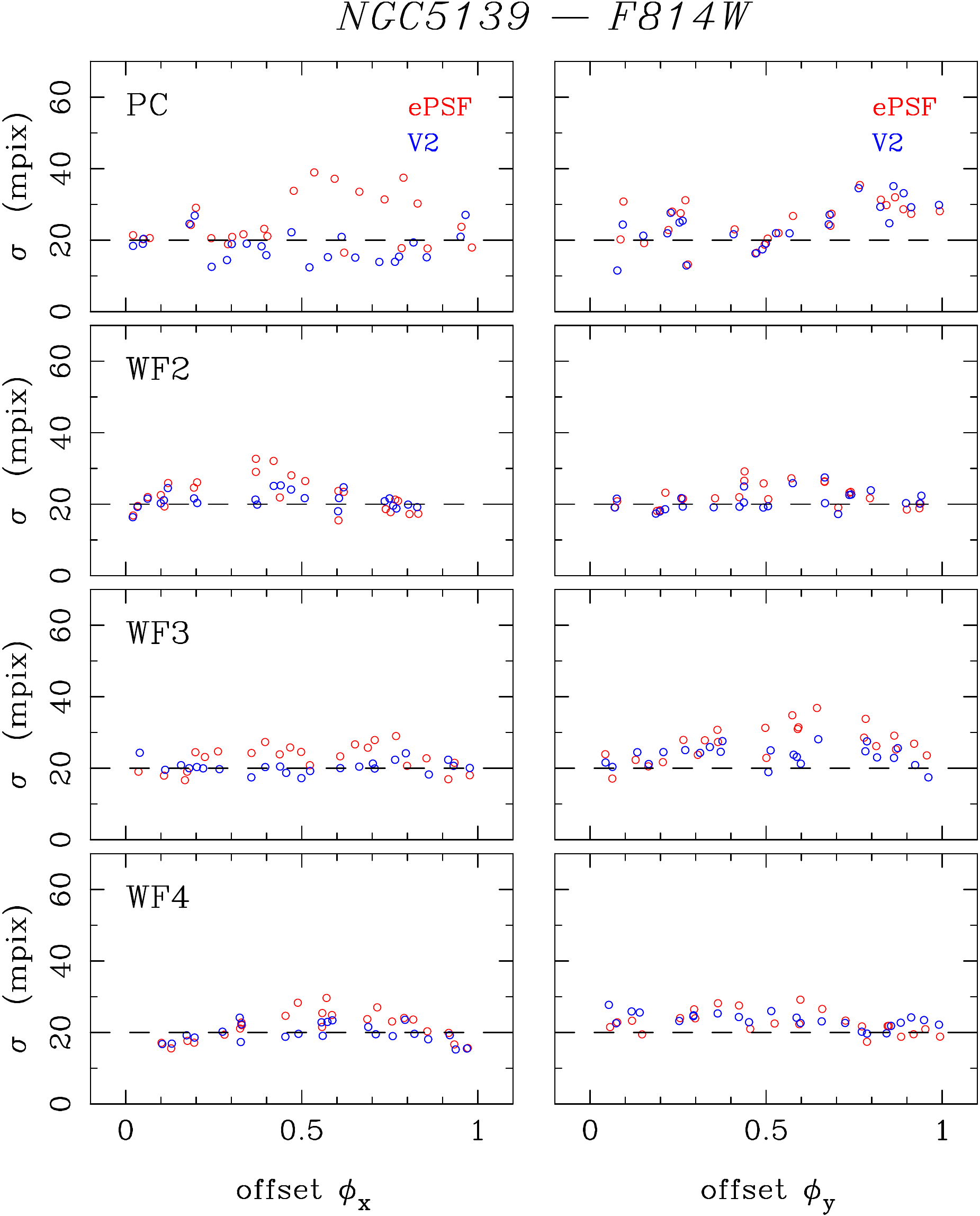}
\caption{As in Fig. \ref{fig:f5-ngc104}, only for NGC 5139 in F814W.}
\label{fig:f8-ngc5139}
\end{figure}

Also in filter F814W, we present results
for three other clusters. The NGC 6656 and
NGC 6205 data sets were taken at about the same
epoch as that of NGC 104, (upon which the DL model is based), while the NGC 5139
set is at a much later epoch (see Tab.
\ref{tab:data-sets}).
Figures \ref{fig:f8-ngc6656}, \ref{fig:f8-ngc6205}
and \ref{fig:f8-ngc5139} present results
for these clusters' data sets. For the PC,
there is little to no improvement of the VGG6
positions compared to the hst1pass ones.
However, for the WF chip, there is notable
improvement seen in Figs. \ref{fig:f8-ngc6656} and
\ref{fig:f8-ngc6205}. Taken at about
the same epoch as the NGC 104 set, it is reasonable that
the PSF for these clusters' data sets 
would be similar to that for the NGC 104 set.
Presumably, it is because of this that
the VGG6 model works reasonably well
correcting the bias in these two sets. 

For the later epoch
NGC 5139 set (Fig. \ref{fig:f8-ngc5139} ),
there is only marginal improvement. This would be in keeping with the
supposition that because of the large epoch difference,
the VGG6-derived model is no longer appropriate.
The PC results for NGC 5139 are also rather
noisy and thus inconclusive. This is because
only of the order of 25 stars were available for
each PC solution, rather than the typical
number of the order of 100 to a few hundred.


\subsection{Per-star Comparison Examples \label{subsec:improve}}

Figures \ref{fig:f5-ngc104} through \ref{fig:f8-ngc5139} represent one way
of illustrating the precision of DL centering compared to that of
ePSF/hst1pass, one in which pixel phase error is highlighted.
Another means of comparison can be made that more directly mimics how
stellar centers are employed in practice, e.g., in proper-motion studies.
How do the standard errors per exposure shown in the aforementioned figures
translate into positional precision for individual stars when averaged over
multiple exposures at a given epoch?
To answer this question we select three exemplary data sets and apply the
following procedure.

\begin{figure} 
\includegraphics[scale=0.47,angle=0]{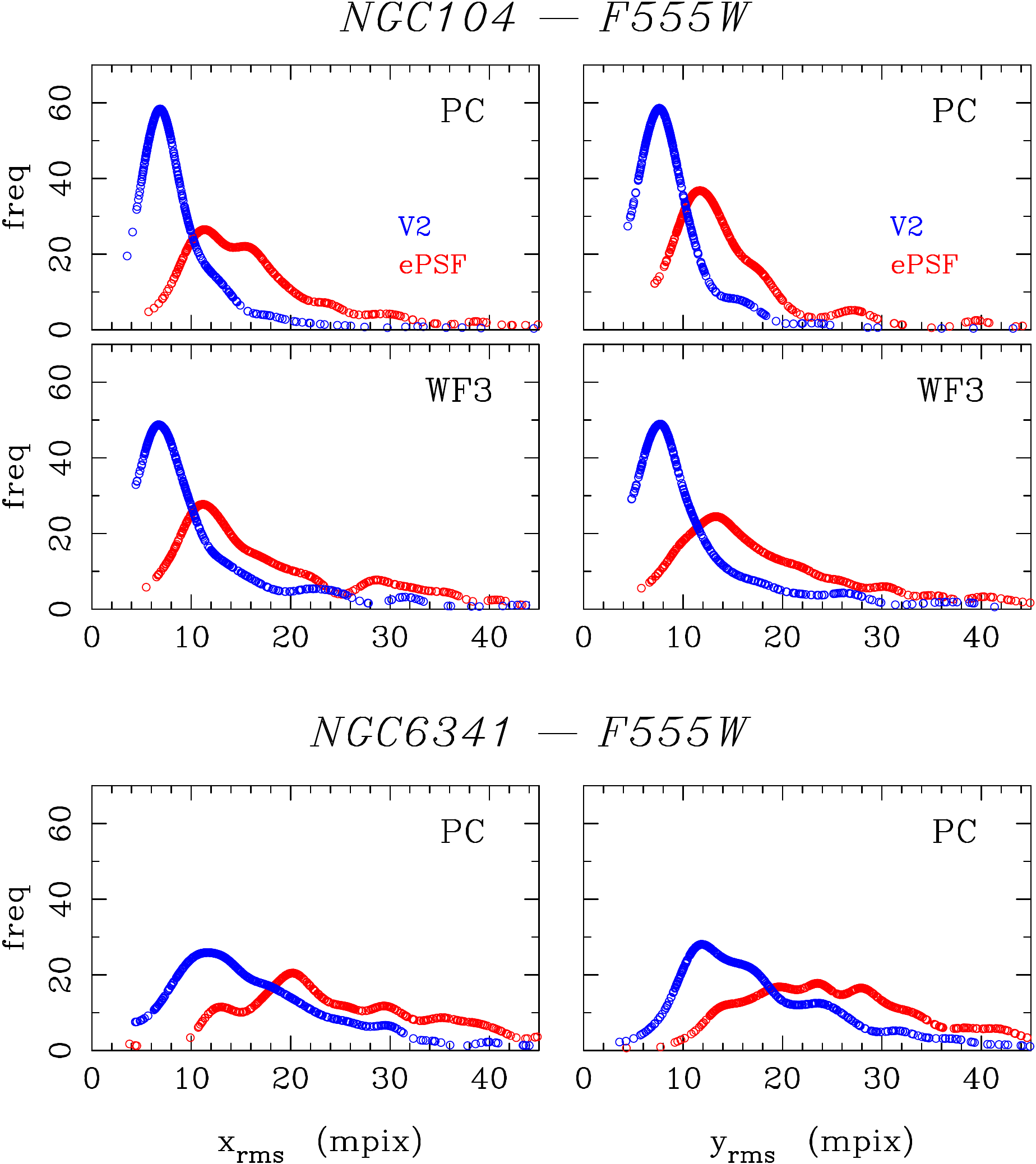}
\caption{Distributions of rms about mean stellar positions
  as derived from repeated measures, based on
  hst1pass/ePSF centers (red symbols) and 
  VGG6-V2 centers (blue symbols).  Distributions along both
  x and y axes are shown for NGC 104
  (PC and WF3) and for NGC 6341 (PC), all in F555W.
  The distributions from hst1pass are broader and shifted
  to higher values compared to those of VGG-V2.
  These can be compared to the range in standard deviation
  seen in the corresponding solutions of these same observations, 
  Figs \ref{fig:f5-ngc104} and \ref{fig:f5-ngc6341}.}
\label{fig:rms-test}
\end{figure}

Star positions for each exposure are transformed into the system of a chosen
reference exposure (just as in Figures \ref{fig:f5-ngc104} through 
\ref{fig:f8-ngc5139}).
Once on a common system, repeat measures of each star are averaged,
discarding any egregious outliers via a conservative sigma clipping.
The resulting rms about each star's mean position is an indicator of its 
astrometric precision.
Distributions of the rms values for the ensemble of stars can then be used
to directly compare the different centering methods.
Figure \ref{fig:rms-test} shows such rms distributions for three data sets:
NGC 104 in F555W, both PC and WF3; and NGC 6341 in F555W, the PC.
The distributions, as shown, are constructed by summing unit-area Gaussian
functions of an adequately narrow width (1.3 mpix) at the location of each 
star's rms value along x or y.

As can be seen in the figure, the VGG6-V2 centers outperform those
of ePSF/hst1pass.
Averaged over the stars depicted in Fig. \ref{fig:rms-test}, 
the ratio of rms values from the
two methods ranges from 1.4 to 1.8 for these three examples, 
i.e., the ePSF/hst1pass errors are from 40 to $80\%$ larger.
Admittedly, the three specific cases selected here are among those in which
the corresponding standard-error plots already show the DL centers to be
superior.
It should be noted that the VGG6-V2 model was trained on a subset of the
images in the NGC 104 exposures, while
the NGC 6341 data set is completely independent.
Nevertheless, in these selected cases the improvement is substantial, as
can be judged directly by the scatter about the average stellar positions.

\subsection{Two Ancillary Tests \label{subsec:twosan}}
We have performed two more tests involving the DL centering procedure
and think it worthwhile including a description of these here.

First, we explore the possibility of taking our DL models based on {\bf simulated}
exposures and applying them to real WFPC2 data.
The utility of such an approach would be considerable, given the dearth of
available WFPC2 data sets (such as those of NGC 104) that have the required star
density and repeated offset exposures to allow for the construction of an empirical 
DL model.
The ability to build reliable DL models from simulated images would provide a means
of reducing WFPC2 exposures at any epoch, yielding star centers free of pixel-phas bias.
As a naive test, we have applied to the real (F555W) exposures of NGC 104 
the DL model of Sec. \ref{subsec:simres}, which was based
on our simulations of the F555W NGC 104 data set.
In effect, this is a test of just how well the simulated images mimic real ones.
Alas, the resulting standard-error plots showed the same
``W'' and ``V''-type curves 
(as in Fig. \ref{fig:stderror-hst1pass} for instance)
indicating significant pixel-phase bias. 
Recall that the simulation code relies on the hst1pass library ePSF profiles,
which our other tests have shown are not ideally suited for the NGC 104 data sets.
An alternate means for constructing a faithful WFPC2 ePSF is required.

Second is a test that is of a completely different nature.
We experiment to see whether any
improvement is to be gained by modifying the real-data DL modeling procedure 
of Sec. \ref{subsec:realin}
with a ``second iteration'' of the average catalog used as the `` ground truth''.
In the original implementation, the catalog was constructed from the trimmed mean
of hst1pass positions for well-measured stars over all 636 exposures.
Perhaps a superior average catalog could be built from positions derived by
applying the ``first-iteration'' DL model to all 636 NGC 104 exposures and using
these to form a trimmed mean for each star, thus greatly reducing any remnant of
the original hst1pass positions.
The second-iteration DL model could then be applied to all 636 exposures and
evaluated as before, by plotting standard error versus offset phase
such as in Fig \ref{fig:f5-ngc104}.
Having done so, we find no significant improvement over the
results of the first-iteration DL model.

We thus conclude that critical to improving
DL-derived positions is the quality of the initial input
``truth'' that goes into the model building.
Thus far we find ourselves limited to the data sets of NGC 104;
perhaps a more sophisticated scheme for combining data sets of different targets
but of similar ePSF composition can be developed.
This will have to be the subject of future study.

\section{Summary \label{sec:concl}}

Pixel-phase bias, acknowledged early
in the studies of WFPC2 images \citep{and2000},
remains a frustrating source of systematic error
even when the ePSF fitting method and library specifically constructed to
address the undersampling issues of WFPC2 are used to center star images.

This bias can be difficult to pinpoint, and only
specially-designed data sets are appropriate to
diagnose and characterize it. It's amplitude can
be as large as 50 mpix (i.e., 2.3 mas for the PC, and 5
mas for the WF) even when using state-of-the-art,
traditional centering algorithms. Such pixel-phase bias
is propagated into positions as unaccounted-for
noise, as subsequent astrometric corrections
(such as the 34th-row and optical distortion corrections)
combined with large offsets between repeated images
will smear out this systematic error and make it appear a random one.

We re-address the elimination of pixel-phase bias by using a novel
approach based on a DL scheme, where
the PSF is learned via training of a VGG6 network 
with simulated images, as well as with real images.

Simulated WFPC2 images helped us develop, test and
characterize the DL model. We obtain
errors of the order of 7-8 mpix for the PC and
6-7 mpix for the WF for typical simulated-star profiles.
High signal-to-noise, very well-measured stars
can achieve 5-mpix errors in the PC
and 3 mpix in the WF. 

Subsequently, we use real WFPC2 images to build another set of DL models.
The unique NGC 104 data set of over 600
exposures, taken with a range of fractional-pixel offsets, is used to
construct an average catalog of positions that is precise and bias free. 
We adopt this catalog as the
``truth'' in the training process for the real images. These DL models,
one for each WFPC2 detector,
are then applied to all detected stars to derive DL-based positions.
These are then compared internally, within the repeated exposures, to test for
pixel-phase bias and astrometric precision, as detailed in
Sec. \ref{subsec:realres}. The DL models derived from the real NGC 104 data set
are applied not only to the NGC 104 data, but also to suitable data sets
of other cluster fields.
For now, we have limited all our analysis to the central portion of each
WFPc2 chip, in order to avoid complications due to a field-varying PSF.
Note that this entire procedure is performed, separately, on real images
in two filters: F555W and F814W.

Doing so, we obtain excellent results for the
NGC 104 data sets. Positional errors are determined from  
the standard errors in Figs. \ref{fig:f5-ngc104}
and \ref{fig:f8-ngc104}, after appropriate division by $\sqrt 2$ since
these standard errors include the measurement error of the
star on the reference image and that on the test image.
The best error values are achieved using version V2 of the VGG6 model.
For the PC we obtain errors of the order of
9 to 10 mpix in filter F555W, and 8.5 mpix in F814W.
For the WF chips we obtain 10 to 11 mpix in F555W, and
8.5 to 9.5 mpix in F814W. 

The DL models built on the NGC 104 data sets are also applied to 
other cluster data sets, as mentioned above.
In these cases we generally obtain significantly improved results over those of the
classic ePSF centering procedure. Exceptions
are those clusters whose observing epoch differed by
many years from that of NGC 104, the set used to
build the DL models. Presumably, the PSF changes with time to an extent
that renders the DL models less of a good fit.
Nonetheless, in every case the DL positions performed at least as well
as the classic ePSF ones.

\section{Future Work \label{sec:future}}
These initial results obtained with
a relatively simple form for the DL model
are very encouraging.
Still, there are some obvious refinements that could be
made to improve the procedure, and these we hope to explore in subsequent
studies.
An immediate objective is to build
a model where the PSF is allowed to vary with position on the chip.
It is known that the variation of the PSF across each chip
is rather mild, \citep{and2000}, therefore
this variation should be easily captured by a straightforward adjustment of the
DL model. Note that the slight increase in model complexity will be more than
compensated for by the increase in the number of stars per chip available to train
the model, a factor of 9 more stars than in our present tests, once the entire
chip is modeled.

Next, we plan to automatically search the hyperparameters
of the DL model to further refine the training process.
More imporantly, we plan to explore other alternatives such as
Residual Networks architectures \citep{he16} with skip-connections,
which allow building deeper neural networks, or Symbolic Regression
\citep{Schmidt2009} to infere analytical expressions from the observed data.


Finally, we will strive to incorporate more WFPC2 data sets into future
model building, as well improve the  of the simulated images.

\acknowledgments

This work was supported by the NASA Connecticut Space
Grant Consortium faculty research
grant 80NSSC20M0129, and by program HST-AR-15632
provided by NASA through a grant from
the Space Telescope Science Institute,
which is operated by the Association of
Universities for Research in Astronomy, Inc.
RBG was funded by the Universidad Internacional de la Rioja (UNIR)
Research Project "ADELA: Aplicaciones de Deep Learning para Astrofísica",
no. B0036, and by the Call for
Grants for Research Stays Abroad 2021/2022 of UNIR.

We thank the referee for helpful suggestions to
improve this manuscript.

We are very grateful to Jay Anderson for his
ongoing support regarding the hst1pass code.
DC thanks her former colleagues, John Onofrey and
Xenios Papademetris at the Radiology and Biomedical
Imaging department of the Yale Medical School, for
introducing her to the DL technique.

\vspace{5mm}
\facilities{{\it HST} (WFPC2), MAST}


\bibliography{ms}{}

\end{document}